\documentclass[11pt]{amsart}
\newtheorem{Theorem}{Theorem}[section]
\newtheorem{Proposition}[Theorem]{Proposition}
\newtheorem{Lemma}[Theorem]{Lemma}

\newtheorem{Assumption 2}[Theorem]{Assumption 2}

\numberwithin{equation}{section}

\usepackage{amsmath}
\usepackage{amssymb}
\usepackage{lmodern}
\usepackage{stmaryrd}
\usepackage{ulem}
\usepackage{yfonts}
\usepackage{amsbsy}
\usepackage{pstricks}
\usepackage{graphicx}
\usepackage{pst-xkey}
\usepackage{pst-all,pst-gr3d}
\usepackage[dvips,arrow,matrix]{xy}
\usepackage{gensymb}

\def\k#1{\kern#1em}
\def\Ib#1{{I\kern-.25em#1}}
\def\Ibb#1{{I\kern-.23em#1}}

\def\AA{{\mathbb A}}
\def\CC{{\mathbb C}}
\def\DD{{\rm\Ib D}}
\def\EE{{\mathbb E}}

\def\HH{{\mathbb H}}

\def\RR{{\mathbb{R}}}

\def\vci{\vrule  width.02em height1.47ex depth-.0ex}
\def\11{{\rm\k{.2}\vci\k{-.37}1}}

\def\fin{{\begin{flushright}
\it{Q.E.D.}
\end{flushright}}}

\begin{document}

\address{Universit\'e de Bordeaux, Institut de Math\'ematiques, UMR CNRS 5251, F-33405 Talence Cedex}

\email{bachelot@math.u-bordeaux1.fr}

\title{New Dynamics in the Anti-De Sitter Universe $AdS^5$}

\author{Alain BACHELOT}

\begin{abstract}
This paper deals with the propagation of the gravitational waves in the Poincar\'e patch of the 5-dimensional Anti-de Sitter universe. We construct a large family of unitary dynamics with respect to some high order energies that are conserved and positive. These dynamics are associated with asymptotic conditions on the conformal time-like boundary of the universe. This result does not contradict the statement of Breitenlohner-Freedman that the hamiltonian is essentially self-adjoint in $L^2$ and thus accordingly the dynamics is uniquely determined. The key point is the introduction of a new Hilbert functional framework that contains the massless graviton which is not normalizable in $L^2$. Then the hamiltonian is not essentially self-adjoint in this new space and possesses a lot of different positive self-adjoint extensions. 
\end{abstract}

\maketitle


\pagestyle{myheadings}
\markboth{\centerline{\sc Alain Bachelot}}{\centerline{\sc New Dynamics  in the Anti-De Sitter Universe $AdS^5$}}

\section{Introduction}
The
5-dimensional Anti-de Sitter space-time $AdS^5$ plays a fundamental role in string cosmology (see e.g. \cite{gibbons}, \cite{mannheim}).
An important geometrical framework is the Poincar\'e patch $\mathcal{P} $ of  $AdS^5$, defined by
\begin{equation*}
\mathcal{P}:=\RR_t\times\RR^3_{\mathbf x}\times]0,\infty[_z,\;\;g_{\mu \nu}dx^{\mu}dx^{\nu}=\frac{1}{z^2}\left(dt^2-d\mathbf{x}^2-dz^2\right).
 \label{}
\end{equation*}
$\mathcal{P}$ is a lorentzian manifold and the crucial point is that it is not globally hyperbolic : the conformal boundary $\RR_t\times\RR_{\mathrm x}\times\{z=0\}$ is time-like and the question arises to determine the possible boundary conditions on this horizon, satisfied by the gravitational waves propagating in the bulk $\mathcal{P}$. These fields obey the D'Alembert equation
\begin{equation}
\square_gu=0,\;\;\;\square_gu:=\mid g\mid^{-\frac{1}{2}}\partial_{\mu}\left(\mid g\mid^{\frac{1}{2}}g^{\mu \nu}\partial_{\nu}u\right)
  \label{kgo}
\end{equation}
If we put $\Phi=:z^{-\frac{3}{2}}u$ the equation (\ref{kgo}) in $\mathcal{P}$ takes the very simple form of the free wave equation on the 1+4-dimensional half Minkowski space-time $\RR_t\times\RR^3_{\mathbf x}\times]0,\infty[_z$, pertubed by a singular cartesian potential $\frac{15}{4z^2}$:
\begin{equation}
\left(\partial_t^2-\Delta_{\mathbf{x}}-\partial_z^2+\frac{15}{4z^2}\right)\Phi=0,\;\;in\;\;\RR_t\times\RR^3_{\mathbf x}\times]0,\infty[_z.
  \label{eq}
\end{equation}
In this work, we adress two questions :\\

{\it(i)} Since  $\mathcal{P}$ is not globally hyperbolic, the dynamics is not {\it a priori} well defined without some boundary condition imposed on the time-like horizon $\{z=0\}$. The usual opinion is that such a supplement constraint is not necessary because the Breitenlohner-Freedman condition is satisfied for the gravitational waves (\cite{breit1}, \cite{ishi2} and Appendix of \cite{DADS}), and so the hamiltonian $-\Delta_{\mathbf{x}}-\partial_z^2+\frac{15}{4z^2}$ is essentially self-adjoint on $C^{\infty}_0(\RR^3_{\mathbf x}\times]0,\infty[_z)$ in the Hilbert space $\mathcal{H}$ choosen to be $L^2(\RR^3_{\mathbf x}\times]0,\infty[_z)$. As a consequence there exists a unique dynamics in the functional framework of the fields with finite energy (\cite{ishi2}, \cite{braneg}):
\begin{equation}
\EE(\Phi):=\int_{\RR^3}\int_0^{\infty}\mid\nabla_{t,{\mathbf x},z}\Phi(t,{\mathbf
  x},z)\mid^2+\frac{15}{4z^2}\mid\Phi(t,{\mathbf x},z)\mid^2d{\mathbf x}dz<\infty.
 \label{enerdir}
\end{equation}
In fact this constraint implies an implicit Dirichlet condition on the boundary of the universe, $\Phi(t,\mathrm{x},0)=0$, and these gravitational waves are called {\it Friedrichs solutions}.
Nevertheless this result of uniqueness is not the end of the story because it depends deeply on the choice of the Hilbert space $\mathcal{H}$ (or the choice of the energy $\EE(\Phi)$). In this paper we show that we can perform a rich variety of different unitary dynamics for the gravitational waves by changing the choice of the conserved energy. We construct a Hilbert space $\mathcal{H}$ such that $-\Delta_{\mathbf{x}}-\partial_z^2+\frac{15}{4z^2}$ is {\it not} essentially self-adjoint on $C^{\infty}_0(\RR^3_{\mathbf x}\times]0,\infty[_z)$ and admits many self-adjoint extensions associated with different boundary conditions at $z=0$ of asymptotic type.\\

{\it(ii)} Another belief is that this cosmological model with a time-like horizon is not physically realistic since the massless graviton $\Phi_G(t,\mathbf{x},z):=z^{-\frac{3}{2}}\phi(t,\mathbf x)$ where $\partial_t^2\phi-\Delta_{\mathbf x}\phi=0$, is not normalizable (in the sense of the $L^2$ norm). In this paper we prove there exists an infinity of pairwise different unitary dynamics for which this graviton is normalizable (in the sense of the new Hilbert space). Moreover these dynamics are not trivial, {\it i.e} any field localized far from $z=0$ at time $t=0$, interacts with the massless graviton : when the field hits the boundary $z=0$, a part of the scattered field is given by the graviton. Furthermore, many of these dynamics are stable in the sense that there is no growing mode and the conserved energy is positive.\\

Now we describe the very simple idea of the construction of these new dynamics. We can see that $\Phi$ is solution of (\ref{eq}) iff $\Psi(t,{\mathbf x},Z):=\mid Z\mid^{-\frac{5}{2}}\Phi(t,{\mathbf x},\mid Z\mid)$ is solution of
\begin{equation}
\left(\partial_t^2-\Delta_{\mathbf{x}}-\Delta_Z\right)\Psi=0,\;\;in\;\;\RR_t\times\RR^3_{\mathbf x}\times\left(\RR^6_Z\setminus\{Z=0\}\right),
 \label{eqmomo}
\end{equation}
and we have proved in \cite{braneg} that $\Phi$ satisfies (\ref{enerdir}) iff $\Psi$ is solution of the free wave equation in the whole Minkowski space-time $\RR_t\times\RR^9_{{\mathbf x},Z}$. As a consequence, to obtain new dynamics for (\ref{eq}), it is sufficient to construct solutions of (\ref{eqmomo}) that are not free waves in $\RR_t\times\RR^9_{{\mathbf x},Z}$. Therefore we look for some self-adjoint extensions of the Laplace operator $\Delta_{\mathbf{x}}+\Delta_Z$ defined on $C^{\infty}_0\left(\RR^3_{\mathbf x}\times\left(\RR^6_Z\setminus\{Z=0\}\right)\right)$. Since this operator is essentially self-adjoint in $L^2(\RR^9)$, we must consider another Hilbert space and try to give a sense to a perturbation localized on $\RR^3_{\mathbf x}\times\{Z=0\}$. It turns out that there has been recent progress on this question, in particular P. Kurasov in 2009 has studied the super-singular perturbations of the Laplacien \cite{kurasov09}. Taking advantage of these novel advances in spectral analysis, we construct some new dynamics for (\ref{eq}) by considering the formal equation
\begin{equation}
\left(\partial_t^2-\Delta_{\mathbf{x}}-\Delta_Z+c\delta_0(Z)\right)\Psi=0,\;\;in\;\;\RR_t\times\RR^3_{\mathbf x}\times\RR^6_Z.
 \label{eqneuf}
\end{equation}
If $\Phi$ is the sum of a field $\Phi_0$ satisfying (\ref{enerdir}), and of a graviton-like singular field $z^{-\frac{3}{2}}\phi(t,\mathbf x)$, then $\Psi(t,{\mathbf x},Z)=\mid Z\mid^{-\frac{5}{2}}\Phi_0(t,{\mathbf x},\mid Z\mid)+\phi(t,\mathbf x) \mid Z\mid^{-4}
$ and  the meaning of the super singular perturbation $c\delta_0(Z)$ is
$$
c\delta_0(Z)\Psi:=-4\pi^3\phi(t,\mathbf x)\delta_0(Z).
$$
A partial Fourier transform with respect to $\mathbf{x}$ allows to reduce the study of (\ref{eqneuf}) to the investigation of the super-singular perturbations of the Klein-Gordon equation 
$$
\left(\partial_t^2-\Delta_Z+m^2+c\delta_0(Z)\right)u=0,\;\;in\;\;\RR_t\times\RR^6_Z,
$$
that we perform in the next section.\\

Finally we summarize our main result.
We look for the gravitational waves solutions of (\ref{eq}) that have an expansion of the following form
\begin{equation*}
\Phi(t,{\mathbf x},z)=\Phi_r(t,{\mathbf x},z)z^{\frac{5}{2}}+\phi_{-1}(t,{\mathbf x})\chi(z)z^{\frac{5}{2}}
+\phi_0(t,{\mathbf x})\chi(z)z^{\frac{5}{2}}\log z+\phi_{1}(t,{\mathbf x})\chi(z)z^{\frac{1}{2}}+\phi_{2}(t,{\mathbf x})z^{-\frac{3}{2}}
 \label{}
\end{equation*}
where $\chi\in C^{\infty}_0(\RR)$, $\chi(z)=1$ in a neighborhood of $0$ and
$
\Phi_r(t,{\mathbf x},0)=0.
$
The term $\phi_{2}(t,{\mathbf x})z^{-\frac{3}{2}}$ is the part of the wave in the sector of the massless graviton. The behaviour of the field on the boundary of the universe is assumed to be for some $(\alpha_0,\alpha_1,\alpha_2)\in\RR^3$ :
\begin{equation}
\phi_{-1}(t,{\mathbf x})
+\alpha_0\phi_0(t,{\mathbf x})+\alpha_1\phi_{1}(t,{\mathbf x})+\alpha_2\phi_{2}(t,{\mathbf x})=0,\;\;t\in\RR,\;\;{\mathbf x}\in\RR^3.
\label{gloglo}
\end{equation}
For a large family of $\alpha_j$, we are able to construct a Hilbert functional framework for which the Cauchy problem associated with (\ref{eq}) is well-posed. At each time, the boundary constraint (\ref{gloglo}) is satisfied and the graviton part $\phi_2$ is non zero even if the initial data are compactly supported far from the boundary of the universe : hence these waves are not Friedrichs solutions. Moreover there exists a conserved energy. This complicated energy involves the derivatives of third order of the fields. An interesting fact is that  this energy is positive for a continuous set of $\alpha_j$, more precisely when
$$
\alpha_2=0,\;\;0<\alpha_1,\;\;-\frac{1}{2}-\frac{3}{2}\log2<\alpha_0+\frac{1}{2}\log\alpha_1<\frac{1}{4}-\frac{1}{2}\log2-\gamma
$$
where $\gamma$ is the Euler's constant. In this important  case, the massless graviton  $\Phi_G(t,\mathbf{x},z):=z^{-\frac{3}{2}}\phi(t,\mathbf x)$ satisfies the constraint (\ref{gloglo}) since $\alpha_2=0$, and its energy is just the usual energy
$$
\EE(\Phi_G)=c\int_{\RR^3_{\mathbf x}}\vert\nabla_{t,\mathbf{x}}\phi(t,\mathbf{x})\vert^2d{\mathbf x}.
$$
Furthermore, the positivity of the conserved energy assures that there  is no growing mode : we can consider that these new possible dynamics of the gravitational fluctuations are stable.


\section{Super-singular perturbation of the wave equation on $\RR^{1+6}$}
We want to investigate the wave equation on the Minkowski space-time $\RR_t\times\RR^6_Z$ with a supersingular perturbation localized at $Z=0$. More precisely, given $m\geq 0$, we shall consider the abstract Klein-Gordon equation
\begin{equation}
\partial^2_tu+\AA u+m^2u=0,
 \label{eqab}
\end{equation}
where $\AA$ is a densely defined selfadjoint operator on a Hilbert space $\HH_0$ of distributions on $\RR^6$, such that
\begin{equation*}
C^{\infty}_0\left(\RR^6\setminus\{0\}\right)\subset Dom(\AA),\;\;\forall \varphi\in C^{\infty}_0\left(\RR^6\setminus\{0\}\right),\;\;\AA\varphi=-\Delta\varphi.
 \label{}
\end{equation*}
In fact, we choose a very simple point-like interaction at the origin, so for all $u\in Dom(\AA)$ , 
$\AA u$ has the form
\begin{equation}
\AA u=-\Delta u+L(u)\delta_0
 \label{aa}
\end{equation}
where $L$ is a continuous linear form on $\HH_0$, equal to zero on $C^{\infty}_0\left(\RR^6\setminus\{0\}\right)$. This constraint yields a character very singular to the perturbation and the Cauchy problem cannot be solved as usual in a scale of Sobolev spaces : if $u\in \cap_{k=0}^2C^{k}\left(\RR_t; H^{s-k}(\RR^6)\right)$ is solution of (\ref{eqab}) and (\ref{aa})  with $L(u)\neq 0$, then $s<-1$ since $\delta_0\in H^{\sigma}(\RR^6)$ iff $\sigma<-3$. Hence a contradiction appears since $C^{\infty}_0(\RR^6\setminus\{0\})$ is dense in  $H^{s}(\RR^6)$, $s\leq 3$, and as a consequence $L(u)=0$. Therefore we have to introduce some functional spaces, in which $C^{\infty}_0(\RR^6\setminus\{0\})$ is not dense. We want also to recover the static solutions $u_{stat}(t,Z)=\mid Z\mid^{-4}$ for $m=0$, and $u_{stat}(t,Z)=\frac{m^2 K_2(m\mid Z\mid)}{2\mid Z\mid^2}$ when $m>0$ where $K_2$ is the classical modified Bessel function (to see below), that are solution of (\ref{eqab}) and (\ref{aa})  with $L(u_{stat})=-4\pi^3$. On the other hand we know (see Lemma \ref{lemah})  that
$$\frac{ m^2K_2(m\mid Z\mid)}{2\mid Z\mid^{-2}}=\frac{1}{\mid Z\mid^4}-\frac{m^2}{4\mid Z\mid^2}-\frac{m^4}{16}\log\mid Z\mid +O(1),\;\;Z\rightarrow 0.
$$
All theses properties suggest to consider Hilbert spaces of distributions, spanned by $\mid Z\mid^{-4}$, $\mid Z\mid^{-2}$, $\log\mid Z\mid$ and some usual Sobolev spaces. More precisely we take $\chi\in C^{\infty}_0(\RR^6_Z)$ satisfying for some $\rho>0$, $\chi(Z)=1$ when $\mid Z\mid\leq \rho$. We introduce the spaces
\begin{equation}
\HH_k:=\left\{u=v_r+v_1\frac{\chi(Z)}{\mid Z\mid^2}+v_2\frac{\chi(Z)}{\mid Z\mid^4},\;v_r\in H^{k+2}(\RR^6_Z),\;\;v_j\in\CC\right\},\;\;k=-1,0,
 \label{ho}
\end{equation}
\begin{equation}
\HH_k:=\left\{u=V_r+v_0\chi(Z)\log(\mid Z\mid)+v_1\frac{\chi(Z)}{\mid Z\mid^2}+v_2\frac{\chi(Z)}{\mid Z\mid^4},\;V_r\in H^{k+2}(\RR^6_Z),\;\;v_j\in\CC\right\},
\;\;k=1,2,
 \label{hk}
\end{equation}
where $H^m(\RR^6)$ are the usual Sobolev spaces of functions $v\in L^2$ such that $(-\Delta+1)^{\frac{m}{2}}v\in L^2$.
It is clear that these spaces do not depend on the choice of function $\chi$, and given $u$, the coordinates $v_j$, $0\leq j\leq 2$, and $V_r(0)$ when $k=2$, are also independent of $\chi$.
We easily check that in the sense of the distributions on $\RR^6_Z$ we have 
\begin{equation}
\Delta_Z\log(\mid Z\mid)=\frac{4}{\mid Z\mid^2},\;\;\Delta_Z\left(\frac{1}{\mid Z\mid^2}\right)=-\frac{4}{\mid Z\mid^4},\;\;\Delta_Z\left(\frac{1}{\mid Z\mid^4}\right)=-4\pi^3\delta_0(Z).
 \label{delta}
\end{equation}
Since  for any $\epsilon>0$, $\delta_0\in H^{-3-\epsilon}(\RR^6)\setminus  H^{-3}(\RR^6)$, we have
\begin{equation*}
\frac{\chi(Z)}{\mid Z\mid^4}\in  H^{-1-\epsilon}(\RR^6)\setminus  H^{-1}(\RR^6),\;\;\frac{\chi(Z)}{\mid Z\mid^2}\in  H^{1-\epsilon}(\RR^6)\setminus  H^{1}(\RR^6),\;\;
\chi(Z)\log(\mid Z\mid) \in  H^{3-\epsilon}(\RR^6)\setminus  H^{3}(\RR^6).
 \label{}
\end{equation*}
We deduce that $\HH_2\subset\HH_1\subset \HH_0\subset\HH_{-1}\subset L^1_{loc}(\RR^6)$. Now we take two real $\mu_1$, $\mu_2$, such that
\begin{equation}
\mu_j<0,\;\;\mu_1\neq\mu_2,
 \label{muj}
\end{equation}
and we choose on $H^2(\RR^6)$ the norm given by :
\begin{equation*}
\parallel v_r\parallel_{H^2}:=\parallel (-\Delta-\mu_1)^{\frac{1}{2}}(-\Delta-\mu_2)^{\frac{1}{2}} v_r\parallel_{L^2}.
 \label{}
\end{equation*}
The other spaces $H^m$ are endowed with the norm $\parallel v_r\parallel_{H^m}:=\parallel (-\Delta+1)^{\frac{m}{2}}v_r\parallel_{L^2}$. If we put
\begin{equation}
\Vert u\Vert_{\HH_k}:=\left(\parallel v_r\parallel_{H^{k+2}}^2+\sum_{j=1}^2\mid v_j\mid^2\right)^{\frac{1}{2}},\;\;k=-1,0,
 \label{norm}
\end{equation}
\begin{equation}
\Vert u\Vert_{\HH_k}:=\left(\parallel V_r\parallel_{H^{k+2}}^2+\sum_{j=0}^2\mid v_j\mid^2\right)^{\frac{1}{2}},\;\;k=1,2,
 \label{normde}
\end{equation}
we can see that $\Vert.\Vert_{\HH_j}$ is a norm on $\HH_j$ and $(\HH_j,\;\parallel.\parallel_{\HH_j})$ is a  Hilbert space, and $\HH_i$ is dense in $\HH_j$ for $j\leq i$. Since $H^{3+\epsilon}(\RR^6)\subset C^0(\RR^6)$, $V_r(0)$ is well defined for any $u\in\HH_2$. Then given a linear form $q$ on $\CC^4$, we introduce the closed subspace of $\HH_2$
\begin{equation*}
\DD(q):=\left\{u\in \HH_2;\;\;q(V_r(0),v_0,v_1,v_2)=0\right\}.
 \label{}
\end{equation*}
$C^{\infty}_0(\RR^6\setminus\{0\})$ is a subspace of $\DD(q)$. We denote $\mathcal{D}'\left(\RR_t;\DD(q_{\lambda})\right)$ the space of the $\DD(q_{\lambda})$-valued vector distributions on $\RR_t$.  Finally we have to choose the linear form $L$ on $\HH_k$. Since we want that $\AA u$ given by (\ref{aa}) belongs to $L^1_{loc}(\RR^6)$, we note that (\ref{delta}) imposes to take :
\begin{equation*}
L(u)=-4\pi^3 v_2.
 \label{}
\end{equation*}
We emphasize that $u\mapsto L(u)\delta_0$ is a local perturbation since when $u=0$ in a neighborhood of $0$, then $v_2=0$, and so $L(u)\delta_0=0$.


\begin{Theorem}
For all $\mu_1$, $\mu_2$ satisfying (\ref{muj}), there exists a continuous family $(q_{\lambda})_{\lambda\in\RR^3}$ of pairwise different linear forms on $\CC^4$ such that $\DD(q_{\lambda})$ is dense in $\HH_1$, and for any $m\geq 0$, $f\in\HH_1$, $g\in\HH_0$, there exists a unique $u_{\lambda}$ satisfying
\begin{equation}
u_{\lambda}\in C^2\left(\RR_t;\HH_{-1}\right)\cap C^1\left(\RR_t;\HH_{0}\right)\cap C^0\left(\RR_t;\HH_1\right)\cap\mathcal{D}'\left(\RR_t;\DD(q_{\lambda})\right),
 \label{regul}
\end{equation}
\begin{equation}
\partial_t^2u_{\lambda}-\Delta_Zu_{\lambda}+m^2u_{\lambda}+L(u_{\lambda})\delta_0=0,
 \label{equu}
\end{equation}
\begin{equation}
u_{\lambda}(0,Z)=f(Z),\;\;\partial_tu_{\lambda}(0,Z)=g(Z).
 \label{condinit}
\end{equation}

The solution depends continuously of the initial data : there exists $C, K>0$, depending of $\lambda$ but independent on $m$,  such that
\begin{equation}
\parallel u_{\lambda}(t)\parallel_{\HH_1}+m\parallel u_{\lambda}(t)\parallel_{\HH_0}+\parallel \partial_tu_{\lambda}(t)\parallel_{\HH_0}\leq C\left(\parallel f\parallel_{\HH_1}+m\parallel f\parallel_{\HH_0}+\parallel g\parallel_{\HH_0}\right)e^{\left(K-m^2\right)_+\mid t\mid},
 \label{estcon}
\end{equation}
where $x_+=x$ when $x>0$ and $x_+=0$ when $x\leq 0$, and for all $\Theta\in C^{\infty}_0(\RR_t)$ we have :
\begin{equation}
\Vert\int\Theta(t)u_{\lambda}(t)dt\Vert_{\HH_2}\leq C\left(\Vert f\Vert_{\HH_1}+m\Vert f\Vert_{\HH_0}+\Vert g\Vert_{\HH_0}\right)\int\ \left(\mid\Theta(t)\mid+\mid\Theta''(t)\mid\right)e^{(K-m^2)_+\mid t\mid}dt.
 \label{dprimehdeu}
\end{equation}

There exists a conserved energy, i.e. a non trivial, continuous quadratic form $\mathcal{E}_{\lambda}$ defined on $\HH_1\oplus\HH_0$, that satisfies :
\begin{equation}
\forall t\in\RR,\;\;\mathcal{E}_{\lambda}\left(u_{\lambda}(t),\partial_tu_{\lambda}(t)\right)=\mathcal{E}_{\lambda}(f,g).
 \label{enr}
\end{equation}
This energy is not positive definite but $\mathcal{E}_{\lambda}$  is given on $C^{\infty}_0(\RR^6\setminus\{0\})\oplus C^{\infty}_0(\RR^6\setminus\{0\})$ by :
\begin{equation}
\mathcal{E}_{\lambda}(f,g)=\parallel \nabla f\parallel_{H^2}^2+m^2\parallel f\parallel_{H^2}^2+\parallel g\parallel_{H^2}^2.
 \label{enerc}
\end{equation}

The dynamics is non trivial : for all $f$, $g$ in $C^{\infty}_0(\RR^6\setminus\{0\})$, if $f$ and $g$ are  spherically symmetric, then $L(u_{\lambda}(t))\neq 0$ for some time $t$, except if $f=g=0$.

If $\lambda\neq\lambda'$ the dynamics are different : given  two spherically symmetric functions $f$, $g$ in $C^{\infty}_0(\RR^6\setminus\{0\})$, $(f,g)\neq(0,0)$, the solutions $u_{\lambda}$ and $u_{\lambda'}$ of  (\ref{regul}), (\ref{equu}), (\ref{condinit}) are different.

The propagation is causal, i.e.
\begin{equation}
supp(u_{\lambda}(t,.))\subset\{Z;\;\mid Z\mid\leq\mid t\mid\}+\left[supp(f)\cup supp(g)\right].
 \label{cause}
\end{equation}

When $f\in\DD(q_{\lambda})$, $g\in\HH_1$, then $u_{\lambda}$ is a strong solution in the sense that :
\begin{equation}
u_{\lambda}\in C^2\left(\RR_t;\HH_0\right)\cap C^1\left(\RR_t;\HH_{1}\right)\cap C^0\left(\RR_t;\DD(q_{\lambda})\right),
 \label{regull}
\end{equation}
and  there exists $C, K>0$, depending of $\lambda$ but independent on $m$,  such that
\begin{equation}
\parallel u_{\lambda}(t)\parallel_{\HH_2}+m\parallel u_{\lambda}(t)\parallel_{\HH_1}+\parallel \partial_tu_{\lambda}(t)\parallel_{\HH_1}\leq C\left(\parallel f\parallel_{\HH_2}+m\parallel f\parallel_{\HH_1}+\parallel g\parallel_{\HH_1}\right)e^{\left(K-m^2\right)_+\mid t\mid}.
 \label{estconer}
\end{equation}

For all $m\geq 0$, $m\neq\sqrt{-\mu_1}$, $m\neq\sqrt{-\mu_2}$, there exists a smooth surface $\Sigma(m)\subset\RR^3$ such that for any $\lambda\in\Sigma(m)$, the static solutions $u(t,Z)=\mid Z\mid^{-4}$ for $m=0$, and $u(t,Z)=\mid Z\mid^{-2}K_2(m\mid Z\mid)$ when $m>0$, belong to $\DD(q_{\lambda})$.
For all $m\geq 0$ and all $\lambda\in\Sigma(0)$, $u_{\pm}(t,Z):=\frac{e^{\pm imt}}{\mid Z\mid^4}$ is a time periodic strong solution of (\ref{equu}), (\ref{regull}).

\label{theoun}
\end{Theorem}

The family of pairwise different linear forms $q_{\lambda}$ is described below by (\ref{cuve}) and the next theorem.
The terrific expression of this high order energy is given by (\ref{energie}) and (\ref{integrafi}). The strategy of the proof consists in introducing a suitable hermitian product $<,>_0$ on $\HH_0$ such that $\AA$ endowed with $\DD(q)$ as domain, is a densely defined self-adjoint operator. Then the energy is simply
$$
{\mathcal E}_{\lambda}(f,g)= \Vert g\Vert_0^2+\left<\AA f,f\right>_0+m^2\Vert f\vert_0^2.
$$

{\it Proof of Theorem \ref{theoun}.}
It will be convenient to use an alternative definition of the spaces $\HH_k$. We take a third real number $\mu_0$ and we assume that
\begin{equation}
\mu_0<0,\;\;\mu_0\neq\mu_1,\;\;\mu_0\neq\mu_2.
 \label{muzero}
\end{equation}
We introduce the distributions
\begin{equation*}
\Phi_0:=(-\Delta-\mu_0)^{-1}(-\Delta-\mu_1)^{-1}(-\Delta-\mu_2)^{-1}\delta_0 \in H^{3-\epsilon}(\RR^6)\setminus  H^{3}(\RR^6),
 \label{}
\end{equation*}
\begin{equation*}
\varphi_j:=(-\Delta-\mu_j)^{-1}\delta_0 \in  H^{-1-\epsilon}(\RR^6)\setminus  H^{-1}(\RR^6).
 \label{}
\end{equation*}
By the elliptic regularity, all these functions belong to $C^{\infty}\left(\RR^6\setminus\{0\}\right)$, and an explicit calculation give the structure near $Z=0$ :


\begin{Lemma}
$\Phi_0$ and $\varphi_j$ belong to $L^1(\RR^6)$ and can be written as
\begin{equation}
 \varphi_j(Z)=\frac{\chi(Z)}{4\pi^3}\left(\frac{1}{\mid Z\mid^4}+\frac{\mu_j}{4\mid Z\mid^2}-\frac{\mu_j^2}{16}\log(\mid Z\mid)\right)+F_j(Z),
 \label{delfij}
\end{equation}
\begin{equation}
 \Phi_0(Z)=\frac{1}{32\pi^3}\chi(Z)\log(\mid Z\mid)+G_0(Z),
 \label{delfo}
\end{equation}
where $F_j$ and $G_0$  are functions of  $H^4(\RR^6)$, satisfying
\begin{equation}
F_j(0)=\frac{\mu_j^2}{256\pi^3}(4\log 2+3-4\gamma-2\log\mid\mu_j\mid),
 \label{fjo}
\end{equation}
\begin{equation}
G_0(0)=-\frac{4\log 2+3-4\gamma}{128\pi^3}-\frac{\mu_1^2(\mu_2-\mu_0)\log\mid\mu_1\mid
+\mu_2^2(\mu_0-\mu_1)\log\mid\mu_2\mid
+\mu_0^2(\mu_1-\mu_2)\log\mid\mu_0\mid}{64\pi^3(\mu_0-\mu_1)(\mu_1-\mu_2)(\mu_2-\mu_0)},
 \label{go}
\end{equation}
where $\gamma$ is the Euler's constant.

As a consequence, we have the following characterization of spaces $\HH_k$ :
\begin{equation}
\HH_k=\left\{u=u_r+u_1\varphi_1(Z)+u_2\varphi_2(Z),\;u_r\in H^{k+2}(\RR^6_Z),\;\;u_j\in\CC\right\},\;\;k=-1,0,
 \label{h'}
\end{equation}
\begin{equation}
\HH_k=\left\{u=U_r+u_0\Phi_0(Z)+u_1\varphi_1(Z)+u_2\varphi_2(Z),\;U_r\in H^{k+2}(\RR^6_Z),\;\;u_j\in\CC\right\},\;\;k=1,2,
 \label{hde'}
\end{equation}
where the coordinates $u_0$, $u_1$, $u_2$ do not depend on the choice of  $\mu_0$, and the norms
\begin{equation}
\vert u\vert_k:=\left(\parallel u_r\parallel_{H^{k+2}}^2+\sum_{j=1}^2\mid u_j\mid^2\right)^{\frac{1}{2}},\;\;k=-1,0,
 \label{norm'}
\end{equation}
\begin{equation}
\vert u\vert_k:=\left(\parallel U_r\parallel_{H^{k+2}}^2+\sum_{j=0}^2\mid u_j\mid^2\right)^{\frac{1}{2}},\;\;k=1,2,
 \label{normde'}
\end{equation}
are equivalent to the $\Vert.\Vert_{\HH_k}$-norms (\ref{norm}), (\ref{normde}).
 \label{lemah}
\end{Lemma}

{\it Proof of Lemma \ref{lemah}.} We use the Bessel formula that gives the Fourier transform $\hat{f}$ of a spherically symmetric function $f\in L^1(\RR^N)$,
$$
\hat{f}( \zeta):=\int_{\RR^N}e^{-iX.\zeta}f(X)dX=\frac{(2\pi)^{\frac{N}{2}}}{\mid \zeta\mid^{\frac{N}{2}-1}}\int_0^{\infty}J_{\frac{N}{2}-1}(\mid\zeta\mid r)F(r)r^{\frac{N}{2}}dr,\;\;F(\mid X\mid):=f(X),
$$
to get :
$$
\Phi_0(Z)=\frac{1}{8\pi^3\mid Z\mid^2}\int_0^{\infty} J_2(z\mid Z\mid)\frac{z^3}{(z^2-\mu_0) (z^2-\mu_1)(z^2-\mu_2)}dz.
$$
We write
$$
\frac{1}{z^2-\mu_j}=2\int_0^{\infty}e^{-(z^2-\mu_j)t_j^2}t_jdt_j,
$$
to obtain
$$
\Phi_0(Z)=\frac{1}{\pi^3\mid Z\mid^2}\int_0^{\infty}\int_0^{\infty}\int_0^{\infty}e^{\mu_0t_0^2+\mu_1t_1^2+\mu_2t_2^2}\left(\int_0^{\infty} J_2(z\mid Z\mid)e^{-z^2(t_0^2+t_1^2+t_2^2)}z^3dz\right)t_0t_1t_2dt_0dt_1dt_2.
$$
We recall formula (10.22.51) of \cite{nist} :
$$
\int_0^{\infty}J_2(z\mid Z\mid)e^{-z^2p^2}z^3dz=\frac{\mid Z\mid^2}{8p^6}e^{-\frac{\mid Z\mid^2}{4p^2}}.
$$
and by replacing in the previous expression, we deduce that
$$
\Phi_0(Z)=\frac{1}{8\pi^3}\int_0^{\infty}\int_0^{\infty}\int_0^{\infty}e^{\mu_0t_0^2+\mu_1t_1^2+\mu_2t_2^2 -\frac{\mid Z\mid^2}{4(t_0^2+t_1^2+t_2^2)}}\frac{t_0t_1t_2}{(t_0^2+t_1^2+t_2^2)^3}dt_0dt_1dt_2.
$$
We use the spherical coordinates of $\RR^3$, $t_0=\rho\cos\varphi\sin\theta$, $t_1=\rho\sin\varphi\sin\theta$, $t_2=\rho\cos\theta$ to get
\begin{equation*}
\begin{split}
\Phi_0(Z)&=\frac{1}{8\pi^3}\int_0^{\infty}\left(\int_0^{\frac{\pi}{2}}\left(\int_0^{\frac{\pi}{2}}e^{\rho^2\sin^2\theta(\mu_0\cos^2\varphi+\mu_1\sin^2\varphi)}\cos\varphi\sin\varphi d\varphi\right)e^{\mu_2\rho^2\cos^2\theta}\sin^3\theta\cos\theta d\theta\right)e^{-\frac{\mid Z\mid^2}{4\rho^2}}\frac{d\rho}{\rho}\\
&=\frac{1}{16\pi^3(\mu_0-\mu_1)}\int_0^{\infty}\left(\int_0^{\frac{\pi}{2}}e^{\rho^2(\mu_0\sin^2\theta+\mu_2\cos^2\theta)}-e^{\rho^2(\mu_1\sin^2\theta+\mu_2\cos^2\theta)}\cos\theta\sin\theta d\theta\right) e^{-\frac{\mid Z\mid^2}{4\rho^2}}\frac{d\rho}{\rho^3}\\
&=\frac{1}{32\pi^3(\mu_0-\mu_1)(\mu_1-\mu_2)}\int_0^{\infty}e^{\mu_1\rho^2-\frac{\mid Z\mid^2}{4\rho^2}}\frac{d\rho}{\rho^5}
+\frac{1}{32\pi^3(\mu_1-\mu_2)(\mu_2-\mu_0)}\int_0^{\infty}e^{\mu_2\rho^2-\frac{\mid Z\mid^2}{4\rho^2}}\frac{d\rho}{\rho^5}\\
&+\frac{1}{32\pi^3(\mu_2-\mu_0)(\mu_0-\mu_1)}\int_0^{\infty}e^{\mu_0\rho^2-\frac{\mid Z\mid^2}{4\rho^2}}\frac{d\rho}{\rho^5}.
\end{split}
\end{equation*}

We can express the modified Bessel function $K_2$ by formula (10.32.10) of \cite{nist} to get
$$
\int_0^{\infty}e^{\mu_j\rho^2-\frac{\mid Z\mid^2}{4\rho^2}}\frac{d\rho}{\rho^5}=-\frac{8\mu_j}{\mid Z\mid^2}K_2(\sqrt{-\mu_j}\mid Z\mid),
$$
therefore we obtain the expression of $\Phi_0$ :
\begin{equation*}
\begin{split}
\Phi_0(Z)=-\frac{1}{4\pi^3\mid Z\mid^2}&\left[\frac{\mu_1}{(\mu_0-\mu_1)(\mu_1-\mu_2)}K_2(\sqrt{-\mu_1}\mid Z\mid) \right.\\
&+\frac{\mu_2}{(\mu_1-\mu_2)(\mu_2-\mu_0)}K_2(\sqrt{-\mu_2}\mid Z\mid)\\
&\left.+\frac{\mu_0}{(\mu_2-\mu_0)(\mu_0-\mu_1)}K_2(\sqrt{-\mu_0}\mid Z\mid)\right]
\end{split}
 \label{}
\end{equation*}
We directly obtain the expression of $\varphi_j$ with a change of variable in formula (II, 3 ; 20) of \cite{schwartz} :
\begin{equation}
\varphi_j(Z)=-\frac{\mu_j}{8\pi^3\mid Z\mid^2}K_2(\sqrt{-\mu_j}\mid Z\mid).
 \label{fij}
\end{equation}
We know that $K_2(z)$ is an analytic function on the surface of the logarithm, and for $z>0$ we have the following asymptotics (see \cite{nist}, formulae (10.25.3) :
\begin{equation*}
K_2(z)\sim\sqrt{\frac{\pi}{2z}}e^{-z},\;\;z\rightarrow\infty,\;\;K_2(z)\sim\frac{2}{z^2},\;\;z\rightarrow 0^+.
 \label{}
\end{equation*}
We deduce that $\Phi_0$ and $\varphi_j$ are in $L^1(\RR^6)$. To derive the asymptotic forms near zero, we use formula (10.31.1) of \cite{nist} that allows to establish that
for $z>0$ :
\begin{equation}
K_2(z)=\frac{2}{z^2}-\frac{1}{2}-\frac{z^2}{8}\log z+z^2F(z^2)+z^4G(z^2)\log z
 \label{delk}
\end{equation}
where $F$ and $G$ are entire and if $\gamma$ denotes the Euler's constant, we have :
\begin{equation}
F(0)=\frac{4\log 2+3-4\gamma}{32}.
 \label{F}
\end{equation}
(\ref{delfij}) follows from (\ref{fij}) and (\ref{delk}) with
\begin{equation*}
\begin{split}
F_j(Z)&=(1-\chi(Z))\varphi_j(Z)\\
&+\chi(Z)\left(-\frac{\mu_j^2}{128\pi^3}\log(-\mu_j)+\frac{\mu_j^2}{8\pi^3}F(-\mu_j\mid Z\mid^2)-\frac{\mu_j^3\mid Z\mid^2}{8\pi^3}G(-\mu_j\mid Z\mid^2)\log(-\mu_j\mid Z\mid)\right).
\end{split}
 \label{}
\end{equation*}
Since $(1-\chi)\varphi_j\in H^{\infty}(\RR^6)$ by elliptic regularity and $\mid Z\mid^2\log(\mid Z\mid)\in H^4_{loc}(\RR^6)$ we conclude that $F_j\in H^4(\RR^6)$, and (\ref{F}) gives (\ref{fjo}). Finally we have
$$
\Phi_0=2\left(\frac{\varphi_1}{(\mu_0-\mu_1)(\mu_1-\mu_2)}+\frac{\varphi_2}{(\mu_1-\mu_2)(\mu_2-\mu_0)}+\frac{\varphi_0}{(\mu_2-\mu_0)(\mu_0-\mu_1)}\right),
$$
hence (\ref{delfo}) follows from (\ref{delfij}) with
\begin{equation*}
G_0=2\left(\frac{F_1}{(\mu_0-\mu_1)(\mu_1-\mu_2)}+\frac{F_2}{(\mu_1-\mu_2)(\mu_2-\mu_0)}+\frac{F_0}{(\mu_2-\mu_0)(\mu_0-\mu_1)}\right),
 \label{}
\end{equation*}
and with this expression of $G_0$, (\ref{go}) follows from (\ref{fjo}). At last the link between $(U_r, u_0, u_1, u_2)$ and $(V_r, v_0, v_1, v_2)$ is easily deduced from 
(\ref{delfij}) and (\ref{delfo}) {\it via} some tedious computations :
\begin{equation}
\begin{split}
v_2=&\frac{1}{4\pi^3}(u_1+u_2),\;\;\;
v_1=\frac{1}{16\pi^3}(\mu_1u_1+\mu_2u_2),\\
v_0=&\frac{1}{64\pi^3}(2u_0-\mu_1^2u_1-\mu_2^2u_2),\;\;\;
V_r=U_r+u_0G_0+u_1F_1+u_2F_2,
\end{split}
 \label{vu}
\end{equation}
\begin{equation}
\begin{split}
u_1=&\frac{16\pi^3v_1-4\pi^3\mu_2v_2}{\mu_1-\mu_2},\;\;u_2=\frac{16\pi^3v_1-4\pi^3\mu_1v_2}{\mu_2-\mu_1},\;\;
u_0=32\pi^3v_0+8\pi^3(\mu_1+\mu_2)v_1-2\pi^3\mu_1\mu_2v_2,\\
u_r=&v_r+\frac{4v_1(\mu_1+\mu_2)-v_2\mu_1\mu_2}{16}\chi(Z)\log(\mid Z\mid)-\frac{16\pi^3v_1-4\pi^3\mu_2v_2}{\mu_1-\mu_2}F_1-\frac{16\pi^3v_1-4\pi^3\mu_1v_2}{\mu_2-\mu_1}F_2,\\
U_r=&V_r-\left(32\pi^3v_0+8\pi^3(\mu_1+\mu_2)v_1-2\pi^3\mu_1\mu_2v_2\right)G_0-\frac{16\pi^3v_1-4\pi^3\mu_2v_2}{\mu_1-\mu_2}F_1-\frac{16\pi^3v_1-4\pi^3\mu_1v_2}{\mu_2-\mu_1}F_2.
\end{split}
 \label{uv}
\end{equation}
These expressions show that the coordinates $u_1$, $u_2$, $u_0$ depend on $u$, $\mu_1$, $\mu_2$, but are independent of the choice of $\mu_0$. Furthermore, since $\chi(Z)\log(\mid Z\mid)\in H^{3-\epsilon}(\RR^6)$ and $F_j,\,G_0\in H^4(\RR^6)$, we see that the $\Vert.\Vert_{\HH_k}$-norms and the $\vert.\vert_k$-norms are equivalent.
\fin


Now we take two real numbers $\gamma_j>0$ and for $u\in \HH_0$ we put
\begin{equation}
\parallel u\parallel_0:=\left(\parallel u_r\parallel_{H^2}^2+\sum_{j=1}^2\gamma_j\mid u_j\mid^2\right)^{\frac{1}{2}},
 \label{normho}
\end{equation}
that is clearly equivalent to the $\Vert.\Vert_{\HH_0}$-norm.
We choose $\theta\in[0,\pi[$ and we put
\begin{equation*}
\lambda:=(\lambda_0,\lambda_1,\lambda_2)=(\cot\theta, \log\gamma_1,\log\gamma_2)\in ]-\infty,\infty]\times\RR\times\RR.
 \label{}
\end{equation*}
We introduce the operator $\AA$ defined by :
\begin{equation*}
\AA u:=-\Delta U_r+\mu_0u_0\Phi_0+\left(\mu_1u_1+\frac{u_0}{\mu_1-\mu_2}\right)\varphi_1+\left(\mu_2u_2+\frac{u_0}{\mu_2-\mu_1}\right)\varphi_2.
 \label{}
\end{equation*}
This operator is a continuous linear map from $\HH_k$ to $\HH_{k-2}$ for $k=1,2$. Now we define $\AA_{\lambda}$ as its restriction to the domain $Dom(\AA_{\lambda})$ defined by
\begin{equation}
Dom(\AA_{\lambda}):=\left\{u\in\HH_2\; ;\;\;U_r(0)\sin\theta+u_0\cos\theta-\left(\gamma_1u_1-\gamma_2u_2\right)\frac{\sin\theta}{\mu_1-\mu_2}=0\right\},
 \label{domal}
\end{equation}
It was proved in \cite{kurasov09} that $(\AA_{\lambda}, Dom(\AA_{\lambda}))$ is a selfadjoint operator on $(\HH_0,\parallel.\parallel_0)$.
We consider the Cauchy problem associated to (\ref{condinit}) and 
\begin{equation}
\partial_t^2u_{\lambda}+\AA_{\lambda}u_{\lambda}+m^2u_{\lambda}=0.
 \label{cocha}
\end{equation}
We show that this equation is just (\ref{equu}). We have $(-\Delta-\mu_0)\Phi_0=(-\Delta-\mu_1)^{-1}\varphi_2=\frac{\varphi_1-\varphi_2}{\mu_1-\mu_2}$. Hence we get
\begin{equation}
\AA u=-\Delta u-(u_1+u_2)\delta_0.
 \label{agagaga}
\end{equation}
Since (\ref{uv}) implies $u_1+u_2=4\pi^3v_2=-L(u)$, the equations (\ref{cocha}) and (\ref{equu}) are equivalent to
 \begin{equation}
\partial_t^2u-\Delta u+m^2u+L(u)\delta_0=0.
 \label{eqf}
\end{equation}
The Cauchy problem for this equation has to be completed by the ``boundary condition at $Z=0$'' specified by the domain of $\AA_{\lambda}$ :
\begin{equation}
U_r(0)\sin\theta+u_0\cos\theta-\left(\gamma_1u_1-\gamma_2u_2\right)\frac{\sin\theta}{\mu_1-\mu_2}=0.
 \label{qlambda}
\end{equation}
Thanks to (\ref{uv}), this constraint can be associated with a linear form $q_{\lambda}\left(V_r(0), v_0, v_1, v_2\right)$ defined on $\CC^4$ and $\DD(q_{\lambda})=Dom(\AA_{\lambda})$. Therefore to prove the Theorem, it is sufficient to investigate the Cauchy problem (\ref{condinit}), (\ref{cocha}).\\

The case $\theta=0$ that corresponds to $u_0=0$ or $ 16v_0+4(\mu_1+\mu_2)v_1-\mu_1\mu_2v_2=0$, is rather peculiar since $Dom(\AA_{\lambda})$ is not dense in $\HH_1$. It corresponds simply to the operator defined on $H^4(\RR^6)\oplus\CC\varphi_1\oplus\CC\varphi_2$ by
\begin{equation}
\forall U_r\in H^4(\RR^6),\;\;\AA_0U_r:=-\Delta U_r,\;\;\AA_0\varphi_j=\mu_j\varphi_j,\;\;j=1,2.
 \label{azero}
\end{equation}
In this case the dynamics is uncoupled between the regular and singular parts of the field : given $f=F_r+f_1\varphi_1+f_2\varphi_2$, $g=G_r+g_1\varphi_1+g_2\varphi_2$, $F_r,G_r\in H^4(\RR^6)$, $f_j,g_j\in\CC$, the Cauchy problem is easily solved by
$$
u_{\lambda}(t,Z)=U_r(t,Z)+f_1(t)\varphi_1(Z)+f_2(t)\varphi_2(Z)
$$
where $U_r$ is the solution of the free Klein-Gordon equation $\partial_t^2U_r-\Delta U_r+m^2U_r=0$ with $U_r(0)=F_r$, $\partial_tU_r(0)=G_r$, and $f_j(t)$ is solution of the harmonic oscillator $\ddot{f_j}+(m^2+\mu_j)f_j=0$, with $f_j(0)=f_j$, $\dot{f_j}(0)=g_j$.\\

In the sequel, we consider the case $\theta\neq 0$, i.e. $\lambda\in\RR^3$ and the family of linear forms is given by
\begin{equation}
q_{\lambda}(V_r(0),v_0,v_1,v_2):=U_r(0)+\lambda_0u_0-\frac{e^{\lambda_1}}{\mu_1-\mu_2}u_1-\frac{e^{\lambda_2}}{\mu_2-\mu_1}u_2.
 \label{qvu}
\end{equation}

First we prove that $Dom(\AA_{\lambda})$ is dense in $\HH_1$. Given $u=U_r+u_0\Phi_0+u_1\varphi_1+u_2\varphi_2\in\HH_1$, we pick a sequence $\psi^n\in C^{\infty}_0(\RR^6\setminus\{0\})$ converging to $U_r$ in $H^3(\RR^6)$, and  a sequence $\chi_n\in C^{\infty}_0(\RR^6\setminus\{0\})$ converging to $\chi$ in $H^3(\RR^6)$.
We put $U_r^n:=\psi_n+\left(\frac{\gamma_1\mu_1-\gamma_2\mu_2}{\mu_1-\mu_2}-u_0\cot\theta\right)(\chi-\chi_n)$. Then $u^n:=U_r^n+u_0\Phi_0+u_1\varphi_1+u_2\varphi_2$ belongs to $Dom(\AA_{\lambda})$ and tends to $u$ in $\HH_1$ as $n$ tends to infinity.

Now we investigate the quadratic form associated with the operator $\AA_{\lambda}$. We use the fact that $\left<(-\Delta-\mu_0)U_r,\Phi_0\right>_{H^2}=U_r(0)=\frac{\gamma_1u_1-\gamma_2u_2}{\mu_1-\mu_2}-u_0\cot\theta$ to evaluate :
\begin{equation}
\begin{split}
\left<\AA_{\lambda}u,u\right>_0=&\mu_0\parallel \Phi_0\parallel^2_{H^2}\mid u_0\mid^2+\mu_1\gamma_1\mid u_1\mid^2+\mu_2\gamma_2\mid u_2\mid^2+\frac{\gamma_1}{\mu_1-\mu_2}u_0\overline{u_1}-\frac{\gamma_2}{\mu_1-\mu_2}u_0\overline{u_2}\\
&+\parallel\nabla U_r\parallel^2_{H^2}+\left<(-\Delta-\mu_0)U_r,u_0\Phi_0\right>_{H^2}+2\mu_0\Re\left<U_r,u_0\Phi_0\right>_{H^2}\\
=&(-\mu_0\parallel \Phi_0\parallel^2_{H^2}-\cot\theta)\mid u_0\mid^2+\mu_1\gamma_1\mid u_1\mid^2+\mu_2\gamma_2\mid u_2\mid^2+2\Re\left(\overline{u_0}\frac{\gamma_1 u_1-\gamma_2 u_2}{\mu_1-\mu_2}\right)\\
&+\parallel\nabla U_r\parallel^2_{H^2}-2\mu_0 \parallel U_r\parallel^2_{H^2}+2\mu_0 \parallel U_r+u_0\Phi_0\parallel^2_{H^2}.
\end{split}
 \label{Auu}
\end{equation}
We see that $u\mapsto \left<\AA_{\lambda}u,u\right>_0$ is a continuous sesquilinear form on $Dom(\AA_{\lambda})$ endowed with the $\HH_1$-norm. Moreover for any $M\geq 0$ we have
\begin{equation}
\begin{split}
 \left<\AA_{\lambda}u,u\right>_0+M\parallel u\parallel^2_0\geq& (-\mu_0\parallel \Phi_0\parallel^2_{H^2}-\cot\theta-1+M)\mid u_0\mid^2\\
&+\gamma_1\left(\mu_1-\frac{\gamma_1}{(\mu_1-\mu_2)^2}+M\right)\mid u_1\mid^2 +\gamma_2\left(\mu_2-\frac{\gamma_2}{(\mu_1-\mu_2)^2}+M\right)\mid u_2\mid^2\\
&+\parallel\nabla U_r\parallel^2_{H^2}-2\mu_0 \parallel U_r\parallel^2_{H^2}+(2\mu_0 +M)\parallel U_r+u_0\Phi_0\parallel^2_{H^2}.
\end{split}
 \label{AuuM}
\end{equation}
We deduce that for $M=M_{\lambda}$ large enough, there exists $\alpha>0$ such that for all $u\in Dom(\AA_{\lambda})$,
\begin{equation}
 \left<\AA_{\lambda}u,u\right>_0+M_{\lambda}\parallel u\parallel^2_0\geq \alpha\parallel u\parallel_{\HH_1}^2.
 \label{coer}
\end{equation}
We conclude that $\AA_{\lambda}$ is bounded from below, $\parallel \left(\AA_{\lambda}+M_{\lambda}\right)^{\frac{1}{2}}u\parallel_0$ is a norm equivalent to the $\HH_1$ norm, the domain of the sesquilinear form is just $\HH_1$, and $\left(\AA_{\lambda}+M_{\lambda}\right)^{-\frac{1}{2}}$ is a continuous linear map from $\HH_0$ to $\HH_{1}$. (\ref{coer}) implies also that
$$
\alpha\parallel u\parallel_{\HH_1}\leq \Vert\left(\AA_{\lambda}+M_{\lambda}\right)u\Vert_{\HH_0},
$$
hence
$$
\sum_{j=0}^2\mid u_j\mid\leq\kappa  \Vert\left(\AA_{\lambda}+M_{\lambda}\right)u\Vert_{\HH_0}.
$$
We have also
$$
\Vert U_r\Vert_{H^4}\leq C\Vert(-\Delta+M_{\lambda})U_r\Vert_{H^2}\leq  C\left(\Vert\left(\AA_{\lambda}+M_{\lambda}\right)u\Vert_{\HH_0}+\mid u_0\mu_0\mid\Vert\Phi_0\Vert_{H^2}\right).
$$
Therefore we conclude that there exists $c(\lambda)>0$ such that for all $u\in Dom(\AA_{\lambda})$ we have
\begin{equation}
\Vert u\Vert_{\HH_2}\leq c(\lambda) \Vert\left(\AA_{\lambda}+M_{\lambda}\right)u\Vert_{\HH_0}\leq \frac{1}{c(\lambda)}\Vert u\Vert_{\HH_2}.
 \label{reghdeu}
\end{equation}
Then it is well-known that for $f\in Dom(\AA_{\lambda})$, $g\in \HH_1$, the Cauchy problem (\ref{condinit}), (\ref{cocha}) has a unique solution $u_{\lambda}\in C^2(\RR_t;\HH_0)\cap C^1(\RR_t;\HH_1)\cap C^0(\RR_t;Dom(\AA_{\lambda}))$ and this solution depends continuously on the initial data (see e.g. theorem 7.8, page 114 in \cite{goldstein}). Nevertheless, since we need to carefully control the constants with respect to the mass $m$, we present some details. If $m^2\geq M_{\lambda}$, we have simply $u_{\lambda}(t)=\cos\left(t\sqrt{\AA_{\lambda}+m^2}\right)f+\frac{\sin\left(t\sqrt{\AA_{\lambda}+m^2}\right)}{\sqrt{\AA_{\lambda}+m^2}}g$, hence (\ref{coer}) and (\ref{reghdeu}) imply :
\begin{equation}
\Vert\partial_tu_{\lambda}(t)\Vert_{\HH_k}+\Vert u_{\lambda}(t)\Vert_{\HH_{k+1}}\leq C\left( \Vert f\Vert_{\HH_{k+1}}+\Vert g\Vert_{\HH_k} \right),\;\;k=0,1.
 \label{enerve}
\end{equation}
When $m^2<M_{\lambda}$, we can construct $u_{\lambda}$ by solving the following integral equation thanks to the Picard's iterates :
$$
u_{\lambda}(t)=\cos\left(t\sqrt{\AA_{\lambda}+M_{\lambda}}\right)f+\frac{\sin\left(t\sqrt{\AA_{\lambda}+M_{\lambda}}\right)}{\sqrt{\AA_{\lambda}+M_{\lambda}}}g+(M_{\lambda}-m^2)\int_0^t  \frac{\sin\left((t-s)\sqrt{\AA_{\lambda}+M_{\lambda}}\right)}{\sqrt{\AA_{\lambda}+M_{\lambda}}}u_{\lambda}(s)ds.
$$
The Gronwall lemma gives 
\begin{equation}
\Vert u_{\lambda}(t)\Vert_{\HH_1}+\Vert\partial_t u_{\lambda}(t)\Vert_{\HH_0}\leq C(\lambda)\left(\Vert f\Vert_{\HH_1}+\Vert g\Vert_{\HH_0}\right)e^{\mid t\mid(M_{\lambda}-m^2)},
 \label{estconn}
\end{equation}
and by applying $\AA_{\lambda}+M_{\lambda}$ to the integral equation, using (\ref{reghdeu}) and the Gronwall lemma again, we get 
\begin{equation}
\Vert u_{\lambda}(t)\Vert_{\HH_2}+\Vert\partial_t u_{\lambda}(t)\Vert_{\HH_1}\leq C(\lambda)\left(\Vert f\Vert_{\HH_1}+\Vert g\Vert_{\HH_0}\right)e^{\mid t\mid(M_{\lambda}-m^2)}.
 \label{estconerr}
\end{equation}
Now we have to control $m\Vert u_{\lambda}(t)\Vert_{\HH_k}$, $k=0,1$. We start by noting that the following energy is conserved :
\begin{equation}
\parallel \partial_tu_{\lambda}(t)\parallel_0^2+\left<\AA_{\lambda}u_{\lambda}(t),u_{\lambda}(t)\right>_0+m^2\parallel u_{\lambda}(t)\parallel_0^2=\parallel g\parallel_0^2+\left<\AA_{\lambda}f,f\right>_0+m^2\parallel f\parallel_0^2:={\mathcal E}_{\lambda}(f,g),
 \label{enerj}
\end{equation}
hence (\ref{coer}) and (\ref{estconn}) imply (\ref{estcon}) with $K:=M_{\lambda}+1$ when $m^2\geq K$. Furthermore, we get its expression with (\ref{Auu}) :  given $f=F_r+f_0\Phi_0+f_1\varphi_1+f_2\varphi_2\in\HH_1$, $g=g_r+g_1\varphi_1+g_2\varphi_2\in\HH_0$, $F_r\in H^3(\RR^6)$, $g_r\in H^2(\RR^6)$, $f_j,g_j\in\CC$, we have :
\begin{equation}
\begin{split}
{\mathcal E}_{\lambda}(f,g)=
&\sum_{j=1}^2e^{\lambda_j}\left[(\mu_j+m^2)\mid f_j\mid^2+\mid g_j\mid^2\right]+
\left(-\mu_0\parallel \Phi_0\parallel^2_{H^2}-\lambda_0\right)\mid f_0\mid^2
+2\Re\left(\overline{f_0}\frac{e^{\lambda_1} f_1-e^{\lambda_2} f_2}{\mu_1-\mu_2}\right)\\
&+\parallel (-\Delta-\mu_1)^{\frac{1}{2}}(-\Delta-\mu_2)^{\frac{1}{2}} g_r\parallel_{L^2}^2+(m^2+2\mu_0)\parallel (-\Delta-\mu_1)^{\frac{1}{2}}(-\Delta-\mu_2)^{\frac{1}{2}} ( F_r+f_0\Phi_0)\parallel_{L^2}^2 \\
&+\parallel\nabla (-\Delta-\mu_1)^{\frac{1}{2}}(-\Delta-\mu_2)^{\frac{1}{2}} F_r\parallel^2_{L^2}-2\mu_0 \parallel (-\Delta-\mu_1)^{\frac{1}{2}}(-\Delta-\mu_2)^{\frac{1}{2}}  F_r\parallel^2_{L^2}\\
&=\sum_{j=1}^2e^{\lambda_j}\left[(\mu_j+m^2)\mid f_j\mid^2+\mid g_j\mid^2\right]+
\left((m^2+\mu_0)\parallel \Phi_0\parallel^2_{H^2}-\lambda_0\right)\mid f_0\mid^2
+2\Re\left(\overline{f_0}\frac{e^{\lambda_1} f_1-e^{\lambda_2} f_2}{\mu_1-\mu_2}\right)\\
&+\parallel (-\Delta-\mu_1)^{\frac{1}{2}}(-\Delta-\mu_2)^{\frac{1}{2}} g_r\parallel_{L^2}^2+2(m^2+2\mu_0)\Re\left(\overline{f_0}(-\Delta-\mu_0)^{-1}F_r(0)\right)  \\
&+\parallel\nabla (-\Delta-\mu_1)^{\frac{1}{2}}(-\Delta-\mu_2)^{\frac{1}{2}} F_r\parallel^2_{L^2}+m^2 \parallel (-\Delta-\mu_1)^{\frac{1}{2}}(-\Delta-\mu_2)^{\frac{1}{2}}  F_r\parallel^2_{L^2},
\end{split}
 \label{energie}
\end{equation}
where we can compute
\begin{equation}
\begin{split}
\parallel \Phi_0\parallel^2_{H^2}=\frac{1}{8}&\int_0^{\infty}\frac{\rho^5}{(\rho^2-\mu_1)(\rho^2-\mu_2)(\rho^2-\mu_0)^2}d\rho\\
=\frac{1}{16}&\left(\frac{\mu_1^2\log(-\mu_1)}{(\mu_2-\mu_1)(\mu_1-\mu_0)^2}
+\frac{\mu_2^2\log(-\mu_2)}{(\mu_1-\mu_2)(\mu_2-\mu_0)^2}\right.\\
&\left.+\frac{(\mu_1\mu_0^2+\mu_2\mu_0^2-2\mu_0\mu_1\mu_2)\log(-\mu_0)}{(\mu_1-\mu_0)^2(\mu_2-\mu_0)^2}
-\frac{\mu_0}{(\mu_1-\mu_0)(\mu_2-\mu_0)}\right).
\end{split}
 \label{integrafi}
\end{equation}
When $f_0=f_1=f_2=g_1=g_2=0$, in particular when $f,g\in C^{\infty}_0(\RR^6\setminus\{0\})$, then ${\mathcal E}_{\lambda}(f,g)$ is given by (\ref{enerc}). To prove (\ref{estconer}) when $m^2\geq M_{\lambda}+1$, we consider for $h\neq 0$, $v_h(t):=h^{-1}[u_{\lambda}(t+h)-u_{\lambda}(t)]$ that tends to $\partial_tu_{\lambda}(t)$ in $C^0(\RR_t;\HH_1)\cap C^1(\RR_t;\HH_0)$ as $h\rightarrow 0$. We apply estimate (\ref{enerj}) to $v_h$ and we get
\begin{equation*}
\begin{split}
\Vert \partial_tv_h(t) \Vert_0^2+&\left \Vert\left(\AA_{\lambda}+M_{\lambda}\right)^{\frac{1}{2}}v_h\right \Vert_0^2+(m^2-M_{\lambda}) \Vert v_h(t) \Vert_0^2\\
=&\left \Vert \frac{\partial_tu_{\lambda}(h)-g}{h}\right \Vert_0^2+\left \Vert\left(\AA_{\lambda}+M_{\lambda}\right)^{\frac{1}{2}}\left(\frac{u_{\lambda}(h)-f}{h}\right)\right \Vert_0^2+(m^2-M)\left \Vert \frac{u_{\lambda}(h)-f}{h}\right \Vert_0^2,
\end{split}
 \label{}
\end{equation*}
and taking the limit as $h$ tends to zero we obtain
\begin{equation*}
\begin{split}
\left\Vert \left(\AA_{\lambda}+M_{\lambda}\right)u_{\lambda}(t)+(m^2-M_{\lambda})u_{\lambda}(t) \right\Vert_0^2+&\left \Vert\left(\AA_{\lambda}+M_{\lambda}\right)^{\frac{1}{2}}\partial_tu_{\lambda}(t)\right \Vert_0^2+(m^2-M_{\lambda}) \Vert \partial_tu_{\lambda}(t)(t) \Vert_0^2\\
=&\left \Vert\AA_{\lambda}f+m^2f\right \Vert_0^2+\left \Vert\left(\AA_{\lambda}+M_{\lambda}\right)^{\frac{1}{2}}g\right \Vert_0^2+(m^2-M)\left \Vert g\right \Vert_0^2.
\end{split}
 \label{}
\end{equation*}
We deduce from this equality and with (\ref{estcon}) and (\ref{reghdeu}), that (\ref{estconer}) is satisfied with $K=M_{\lambda}+1$ when $m^2\geq M_{\lambda}+1$. It remains to study the case $0\leq m^2\leq M_{\lambda}+1$. We simply use  (\ref{estconn}) and (\ref{estconerr}) to write
$$
\sup_{m^2\leq M_{\lambda}+1}m\Vert u_{\lambda}(t)\Vert_{\HH_k}\leq K\left( \Vert f\vert_{\HH_k}+\left\vert\int_0^t\Vert\partial_tu_{\lambda}(s)\Vert_{\HH_k}ds\right\vert\right)\leq C'(\lambda)\left(\Vert f\Vert_{\HH_{k+1}}+\Vert g\Vert_{\HH_k}\right)e^{\mid t\mid(K-m^2)}.
$$
Now (\ref{estcon}) and (\ref{estconer}) are straight consequences of this estimate and (\ref{estconn}) and (\ref{estconerr}).\\

To solve the Cauchy problem when $(f,g)\in\HH_1\oplus\HH_0$, we pick a sequence $(f^n,g^n)\in Dom(\AA_{\lambda})\oplus\HH_1$ that tends to $(f,g)$ in $\HH_1\oplus\HH_0$ as $n\rightarrow\infty$. Estimation (\ref{estcon}) assures that the solution $u^n\in C^2(\RR_t;\HH_0)\cap C^1(\RR_t;\HH_1)\cap C^0(\RR_t;Dom(\AA_{\lambda}))$ of the Cauchy problem with initial data $(f^n,g^n)$ tends to a function $u\in C^1(\RR_t;\HH_0)\cap C^0(\RR_t;\HH_1)$ that is solution of (\ref{condinit}), (\ref{cocha}) and satisfies (\ref{enerj}). Since $\AA$ is continuous from $\HH_1$ to $\HH_{-1}$, the equation gives $u\in C^2(\RR_t;\HH_{-1})$. To prove that $u$ is a distribution of $\mathcal{D}'\left(\RR_t;Dom(\AA_{\lambda})\right)$, we take $\Theta\in C^{\infty}_0(\RR_t)$ and we consider $F:=\int u(t)\Theta(t)dt\in \HH_1$ and $F^n:=\int u^n(t)\Theta(t)dt\in Dom(\AA_{\lambda})$. By the previous argument $F^n$ tends to  $F$ in $\HH_1$ as $n\rightarrow\infty$. Moreover $\AA_{\lambda}F^n=-\int u^n(t)({\Theta}''(t)+m^2\Theta(t))dt$ that converges to $-\int u(t)({\Theta}''(t)+m^2\Theta(t))dt$ in $\HH_1$. We conclude with (\ref{reghdeu}) and (\ref{estcon}) that $F\in Dom(\AA_{\lambda})$, i.e. $u$ is a $Dom(\AA_{\lambda})$-valued distribution on $\RR_t$ and (\ref{dprimehdeu}) is established.\\

To prove the uniqueness, we consider a solution $u$ satisfying (\ref{regul}), (\ref{cocha}) and  (\ref{condinit}) with $f=g=0$. We take a test function $\Theta\in C^{\infty}_0(\RR_t)$, $0\leq\Theta$, $\int\Theta(t)dt=1$, and we define $u^n(t)=n\int\Theta(ns)u(t+s)ds$.
 $u^n$ tends to $u$ in $C^1(\RR_t;\HH_0)\cap C^0(\RR_t;\HH_1)$ as $n\rightarrow\infty$, hence we have $\parallel u^n(0)\parallel_{\HH_1}\rightarrow 0$, $\parallel \partial_tu^n(0)\parallel_{\HH_0}\rightarrow 0$.
Moreover $u^n$ is a strong solution satisfying (\ref{regull}) and (\ref{estcon}). Therefore $u^n$ tend to $0$ in $C^1(\RR_t;\HH_0)\cap C^0(\RR_t;\HH_1)$, and finally $u=0$.\\

We prove now  that the propagation is causal.  We write $u_{\lambda}=W+w$ where $W$ is solution of the free Klein-Gordon equation $(\partial_t^2-\Delta+m^2)W=0$ with $W(0)=u_{\lambda}(0)$, $\partial_tW(0)=\partial_tu_{\lambda}(0)$. Then $(\partial_t^2-\Delta+m^2)w=-L(u_{\lambda})\delta_0$ with $w(0)=\partial_tw(0)=0$. We have $supp(W(t,.))\subset\{Z;\;\mid Z\mid\leq\mid t\mid\}+\left[supp(f)\cup supp(g)\right]$, $supp(w(t,.))\subset\{Z;\;\mid Z\mid\leq\mid t\mid\}$. When $0\in supp(f)\cup supp(g)$, $supp(w(t,.))\subset supp(W(t,.))$ and (\ref{cause}) is established. When $0\notin supp(f)\cup supp(g)$, we consider firstly the case $(f,g)\in Dom(\AA_{\lambda})\oplus\HH_1$. then necessarily $u_0(0)=\dot{u}_0(0)=u_1(0)=\dot{u}_1(0)=u_2(0)=\dot{u}_2=0$ hence $(f,g)\in H^4(\RR^6)\times H^3(\RR^6)$. We denote $\tau>0$ the distance between $0$ and $supp(f)\cup supp(g)$. For $\mid t\mid\leq\tau$, $W(t)$ satisfies trivially the boundary constraint $q_{\lambda}(W(t))=0$, hence $W(t)=u_{\lambda}(t)$. As a consequence $L(u(t))=0$ for $\mid t\mid\leq\tau$, and for all $t$, $supp(w(t))\subset\{Z;\;\;\mid Z\mid\leq \mid t\mid-\tau\}$. Since $0\in\{Z;\;\mid Z\mid\leq\mid \tau\mid\}+\left[supp(f)\cup supp(g)\right]$, we conclude that (\ref{cause}) is satisfied again. When $(f,g)\in H^3(\RR^6)\oplus H^2(\RR^6)$ and $0\notin supp(f)\cup supp(g)$, we choose a sequence $(f^n,g^n)\in H^4(\RR^6)\times H^3(\RR^6)$ that tends to $(f,g)$ in $H^3(\RR^6)\oplus H^2(\RR^6)$, and $supp(f^n)\cup supp(g^n)\subset\{Z;\;\mid Z\mid\leq\frac{1}{n}\}+\left[supp(f)\cup supp(g)\right]$. The previous result assures that $supp(u_{\lambda}^n(t,.))\subset\{Z;\;\mid Z\mid\leq\mid t\mid+\frac{1}{n}\}+\left[supp(f)\cup supp(g)\right]$ where $u_{\lambda}^n$ is the strong solution with initial data $(f^n, g^n)$. Now (\ref{cause}) follows from the convergence of $u_{\lambda}^n$ to $u_{\lambda}$ in $C^0(\RR_t;\HH_1)$ as $n\rightarrow\infty$.\\

To show that the dynamics is not trivial and involves a singular part in $\mid Z\mid^{-4}$ even for smooth initial data, we consider a solution $u_{\lambda}$ with spherically symmetric Cauchy data $f,g\in C^{\infty}_0(\RR^6\setminus\{0\})$, and we assume that $L(u_{\lambda}(t))=0$ for any time $t$. Then $u_{\lambda}$ is a smooth solution of the free Klein-Gordon equation, and so $u_{\lambda}(t,.)\in C^{\infty}_0(\RR^6)$ for all $t$. Therefore $u_0(t)=u_1(t)=u_2(t)=0$ and $u_{\lambda}=U_r$. The constraint $u_{\lambda}\in Dom(\AA_{\lambda})$ implies $U_r(t,0)=0$. Moreover the Fourier transform of $u_{\lambda}$ is given by the classical formula
$$
\mathcal{F}(u_{\lambda})(t,\zeta)=\sum_{\pm}e^{\pm it\sqrt{\mid \zeta\mid^2+m^2}}A_{\pm}(\mid\zeta\mid).
$$
Then $u_{\lambda}(t,0)=0$ implies that for any $t\in\RR$, 
$$
\int_{-\infty}^{\infty}e^{itr}\left(A_+(\sqrt{r^2-m^2})\mathbf{1}_{[m,\infty[}(r)+A_-(\sqrt{r^2-m^2})\mathbf{1}_{]-\infty,-m]}(r)\right)(r^2-m^2)^2rdr=0.
$$
We conclude that $A_{\pm}=0$ and finally $f=g=0$.\\

Now we show that different $\lambda$ yield to different dynamics. We assume that $u=U_r+u_0\Phi_0+u_1\varphi_1+u_2\varphi_2$ is solution of (\ref{equu}), (\ref{condinit}), (\ref{regull}) for some $\lambda$ and $\lambda'$ in $\RR^3$ with spherically symmetric initial data $f,g\in C^{\infty}_0(\RR^6\setminus\{0\})$, $(f,g)\neq (0,0)$. Since $u\in C^2(\RR_t;\HH_0)$, we have $u_r:=U_r+u_0\Phi_0\in C^2(\RR_t;H^2(\RR^6))$, $u_1$ $u_2\in C^2(\RR)$. From $u\in C^1(\RR_t;\HH_1)$ we deduce that $u_0\in C^1(\RR)$ and $U_r\in C^1(\RR_t;H^3(\RR^6))$. Finally $u\in C^0(\RR_t; \DD(q_{\lambda}))$ implies $U_r\in C^0(\RR_t;H^4(\RR^6))$. Furthermore (\ref{equu}) implies that
\begin{equation*}
\partial_t^2u_r+\ddot{u}_1\varphi_1+\ddot{u}_2\varphi_2-\Delta U_r+\mu_0u_0\Phi_0+m^2u_r+\left(\mu_1u_1+m^2u_1+\frac{u_0}{\mu_1-\mu_2}\right)\varphi_1+\left(\mu_2u_2+m^2u_2+\frac{u_0}{\mu_2-\mu_1}\right)\varphi_2=0,
 \label{}
\end{equation*}
where $\ddot{u}_j$ denotes the second derivative in time. By examining the regularity of each term, we obtain :
\begin{equation}
\begin{split}
&\partial_t^2u_r-\Delta U_r+\mu_0u_0\Phi_0+m^2u_r=0,\\
&\ddot{u}_1 +(\mu_1+m^2)u_1+\frac{u_0}{\mu_1-\mu_2}=0,\\
&\ddot{u}_2+(\mu_2+m^2)u_2+\frac{u_0}{\mu_2-\mu_1}=0.
\end{split}
 \label{krotch}
\end{equation}
This system has to be completed by the initial data
\begin{equation*}
U_r(0)=f,\;\;u_0(0)=u_1(0)=u_2(0)=0,\;\;\partial_tU_r(0)=g,\;\;\dot{u}_0(0)=\dot{u}_1(0)=\dot{u}_2(0)=0,
 \label{}
\end{equation*}
and the boundary condition at $Z=0$ :
\begin{equation*}
U_r(t,0)+\lambda_0u_0(t)-\frac{e^{\lambda_1}}{\mu_1-\mu_2}u_1(t)-\frac{e^{\lambda_2}}{\mu_2-\mu_1}u_2(t)=
U_r(t,0)+\lambda_0'u_0(t)-\frac{e^{\lambda_1'}}{\mu_1-\mu_2}u_1(t)-\frac{e^{\lambda_2'}}{\mu_2-\mu_1}u_2(t)=0.
 \label{}
\end{equation*}
We get from these both constraints that
\begin{equation}
(\lambda_0-\lambda_0')u_0(t)=\frac{e^{\lambda_1}-e^{\lambda_1'}}{\mu_1-\mu_2}u_1(t)+\frac{e^{\lambda_2-}e^{\lambda_2'}}{\mu_2-\mu_1}u_2(t).
 \label{gloups}
\end{equation}
First we assume that $\lambda_0\neq\lambda_0'$. Then $u_0=\frac{\lambda_0-\lambda_0'}{\mu_1-\mu_2}\left([e^{\lambda_1}-e^{\lambda_1'}]u_1-[e^{\lambda_2}-e^{\lambda_2'}]u_2\right)$, hence
\begin{equation}
\begin{split}
&\ddot{u}_1 +(\mu_1+m^2)u_1+\frac{\lambda_0-\lambda_0'}{(\mu_1-\mu_2)^2}\left([e^{\lambda_1}-e^{\lambda_1'}]u_1-[e^{\lambda_2}-e^{\lambda_2'}]u_2\right)=0,\\
&\ddot{u}_2+(\mu_2+m^2)u_2-\frac{\lambda_0-\lambda_0'}{(\mu_1-\mu_2)^2}\left([e^{\lambda_1}-e^{\lambda_1'}]u_1-[e^{\lambda_2}-e^{\lambda_2'}]u_2\right)=0.
\end{split}
 \label{qudiff}
\end{equation}
Since the initial data for $u_j$ are zero, we deduce that $u_1(t)=u_2(t)=0$ for all $t$, that is a contradiction with the fact that $u_1+u_2$ is not identically zero. We conclude that $\lambda_0=\lambda_0'$. As a consequence of (\ref{gloups}), we get
\begin{equation*}
\left(e^{\lambda_1}-e^{\lambda_1'}\right)u_1(t)=\left(e^{\lambda_2}-e^{\lambda_2'}\right)u_2(t).
 \label{}
\end{equation*}
We assume that $\lambda_1\neq \lambda_1'$, hence we can express $u_1$ in term of $u_2$. Since (\ref{krotch}) shows that $\ddot{u}_1+\ddot{u}_2+(\mu_1+m^2)u_1+(\mu_2+m^2)u_2=0$, we deduce that $u_2$ is solution of
$$
\left(\frac{e^{\lambda_2}-e^{\lambda_2'}}{e^{\lambda_1}-e^{\lambda_1'}}+1\right)\ddot{u}_2+\left((\mu_1+m^2)\frac{e^{\lambda_2}-e^{\lambda_2'}}{e^{\lambda_1}-e^{\lambda_1'}}+(\mu_2+m^2)\right)u_2=0,\;\;u_2(0)=\dot{u}_2(0)=0.
$$
If $\frac{e^{\lambda_2}-e^{\lambda_2'}}{e^{\lambda_1}-e^{\lambda_1'}}\neq -1$, then $u_2(t)=0$ for all $t$, hence $u_1$ is also zero, that is a contradiction as previous.
If $\frac{e^{\lambda_2}-e^{\lambda_2'}}{e^{\lambda_1}-e^{\lambda_1'}}= -1$, then $(\mu_1-\mu_2)u_2(t)=0$ for all $t$, hence $u_j$ is also zero, that is a contradiction again.
We conclude that $\lambda_1=\lambda_1'$. We can prove by the same way that $\lambda_2=\lambda_2'$, and finally $\lambda=\lambda'$.\\

We now want to determine for which $\lambda$, the static solutions belong to $Dom(\AA_{\lambda})$. Such a solution is given by $u_{stat}:=(-\Delta+m^2)^{-1}\delta_0$ which is equal to $\frac{m^2}{8\pi^3\mid Z\mid^2}K_2(m\mid Z\mid)$ when $m\neq 0$ and $-\frac{1}{4\pi^3\mid Z\mid^4}$ for $m=0$, since $L(u_{stat})=-1$. If we write $u_{stat}=U_r+u_0\Phi_0+u_1\varphi_1+u_2\varphi_2$, we deduce from (\ref{delfij}) and (\ref{delfo}) that $u_{stat}\in\HH_2$, and its coordinates are given by :
\begin{equation*}
u_1=\frac{m^2+\mu_2}{\mu_2-\mu_1},\;\;u_2=\frac{m^2+\mu_1}{\mu_1-\mu_2},\;\;u_0=-\frac{m^4+m^2(\mu_1+\mu_2)+\mu_1\mu_2}{2},
 \label{}
\end{equation*}
\begin{equation*}
U_r(0)=-\frac{m^4}{8\pi^3}\log m+\frac{m^4}{8\pi^3}F(0)-u_0G_0(0)-u_1F_1(0)-u_2F_2(0),
 \label{}
\end{equation*}
where $F(0)$, $G_0(0)$ and $F_j(0)$ are given by (\ref{fjo}), (\ref{go}) and (\ref{F}).
Since $\mu_j\neq-m^2$, then $u_0\neq 0$. Therefore $u_{stat}\in Dom(\AA_{\lambda})$ iff
\begin{equation*}
\lambda\in\Sigma(m):=\left\{\lambda\in\RR^3;\;\;\lambda_0=\frac{1}{u_0}\left(\frac{e^{\lambda_1}}{\mu_1-\mu_2}u_1+ \frac{e^{\lambda_2}}{\mu_2-\mu_1}u_2-U_r(0)\right)\right\}.
 \label{}
\end{equation*}
At last it is clear that the time periodic solution $e^{\pm imt}\mid Z\mid^{-4}$ is in $Dom(\AA_{\lambda})$ iff $\lambda\in\Sigma(0)$.

\fin

The previous construction heavily depends on the choice of the different parameters $\mu_0$, $\mu_1$, $\mu_2$, $\theta$, $\gamma_1$, $\gamma_2$. We now want to make more clear the role of these parameters.  First we note that the changing of $\mu_0$ into $\mu_0'$, does not affect $u_0$, $u_1$, $u_2$ and it reduces to replace
$\lambda_0$ by $\lambda_0+G'_0(0)-G_0(0)$, where $G'_0(0)$ is defined by (\ref{go}) with $\mu_0'$, $\mu_1$, $\mu_2$. Therefore, the set of all the linear forms
\begin{equation*}
q_{\lambda}(V_r(0),v_0,v_1,v_2)=AV_r(0)+\alpha_0v_0+\alpha_1v_1+\alpha_2v_2,\;\;A,\alpha_j\in\RR
 \label{}
\end{equation*}
is obtained by varying $\mu_1$, $\mu_2$, $\mu_1\neq\mu_2$, $\mu_j<0$, $\theta\in\RR$, $\gamma_1,\gamma_2>0$.\\

As we have noticed, the case $\theta=0$ in (\ref{qlambda}) is not very interesting since  in this case, the dynamics is trivial for the initial data $f$, $g$ in $C^{\infty}_0(\RR^6\setminus\{0\})$ : the solution $u$ satisfies $L(u)=0$ and $\partial_t^2u-\Delta u=0$. It corresponds to the condition $u_0=0$ that becomes by (\ref{uv}) with $\mu'_j=\mu_j/4$
$$
A=0,\;\;v_0+(\mu'_1+\mu'_2)v_1-\mu'_1\mu'_2v_2=0,
$$
where $\mu'_j:=\mu_j/4$ are any real numbers such that $\mu'_j<0$, $\mu'_1\neq\mu'_2$. If we put $\alpha_1=\mu'_1+\mu'_2$, $\alpha_2=-\mu'_1\mu'_2$, then $\mu'_j$ are solution of the polynomial $\mu^2-\alpha_1\mu-\alpha_2=0$. This equation has two negative distinct solutions if and only if the coefficients  $\alpha_j$ satisfy
\begin{equation}
\alpha_0=1,\;\;\alpha_1<0,\;\;-\alpha_1^2<4\alpha_2<0,
 \label{atri}
\end{equation}

For $\theta\neq 0$,  we describe  in terms of the coordinates $(V_r(0)$, $v_0$,$ v_1$, $v_2)$, all the families of the linear forms that we have constructed. If we normalize by taking $A=1$,  (\ref{uv}) and (\ref{qvu}) show that :
\begin{equation}
\begin{split}
\alpha_0=&32\pi^3\left(\lambda_0-G_0(0)\right),\\
\alpha_1=&(\mu_1+\mu_2)\left[8\pi^3\left(\lambda_0-G_0(0)\right)-\frac{\log2}{4}-\frac{3}{16}+\frac{\gamma}{4}\right]+\frac{\mu_1^2\log(\mid\mu_1\mid)-\mu_2^2\log(\mid\mu_2\mid)}{8(\mu_1-\mu_2)}-16\pi^3\frac{\gamma_1+\gamma_2}{(\mu_1-\mu_2)^2},\\
\alpha_2=&\mu_1\mu_2\left[-2\pi^3\left(\lambda_0-G_0(0)\right)+\frac{\log2}{16}+\frac{3}{64}-\frac{\gamma}{16}-\frac{\mu_1\log(\mid\mu_1\mid)-\mu_2\log(\mid\mu_2\mid)}{32(\mu_1-\mu_2)}\right]+4\pi^3\frac{\mu_1\gamma_1+\mu_2\gamma_2}{(\mu_1-\mu_2)^2},
\end{split}
 \label{cuve}
\end{equation}
where $G_0(0)$ can be explicitly expressed by the formula (\ref{go}) involving the Euler's constant $\gamma$, and $\mu_0,\mu_1,\mu_2$.

Conversely, we want to determine for which $\alpha:=(\alpha_0,\alpha_1,\alpha_2)$, $V_r(0)+\alpha_0v_0+\alpha_1v_1+\alpha_2v_2$ is a linear form $q_{\lambda}$ associated with some $\mu_j<0$, $\mu_1\neq\mu_2$, and $\gamma_j>0$.

\begin{Theorem}
 \label{family}
The whole family of the linear forms 
$$
q_{\lambda}(V_r(0),v_0,v_1,v_2)=V_r(0)+\alpha_0v_0+\alpha_1v_1+\alpha_2v_2,\;\;\alpha_j\in\RR
$$
of the Theorem \ref{theoun} obtained with all the values of $\mu_1, \mu_2<0$, $\mu_1\neq\mu_2$, $\lambda\in\RR^3$, is given by the set $\mathcal{A}$ of $\alpha\in\RR^3$ satisfying firstly
\begin{equation}
\alpha_0+\frac{\alpha_1}{\alpha_1+\sqrt{\alpha_1^2-4\alpha_2}}-\frac{\alpha_2}{\left(\alpha_1+\sqrt{\alpha_1^2-4\alpha_2}\right)^2}+\frac{1}{2}\log\left\vert\alpha_1+\sqrt{\alpha_1^2-4\alpha_2}\right\vert<\frac{3}{4}-\gamma,
 \label{zin}
\end{equation}
and secondly
\begin{equation}
\left\{
\begin{array}{c}
\alpha_2<0,\\
or\\
\alpha_2=0,\;\;\alpha_1>0,\\
or\\
\left\{\begin{array}{c}
0<\alpha_1,\;\;0<4\alpha_2<\alpha_1^2,\\
\alpha_0+\frac{\alpha_1}{\alpha_1-\sqrt{\alpha_1^2-4\alpha_2}}-\frac{\alpha_2}{\left(\alpha_1-\sqrt{\alpha_1^2-4\alpha_2}\right)^2}+\frac{1}{2}\log\left\vert\alpha_1-\sqrt{\alpha_1^2-4\alpha_2}\right\vert>\frac{3}{4}-\gamma.
\end{array}
\right.
\end{array}
\right.
 \label{zinb}
\end{equation}

If $(\alpha_0,\alpha_1,\alpha_2)\neq (\alpha_0',\alpha_1',\alpha_2')$, the dynamics are different :  given  two spherically symmetric functions $f$, $g$ in $C^{\infty}_0(\RR^6\setminus\{0\})$, $(f,g)\neq(0,0)$, the solutions $u$ and $u'$ of  (\ref{regul}), (\ref{equu}), (\ref{condinit}) are different.
 
Given $m>0$, the static solution $u(Z)=\frac{K_2(m\mid Z\mid)}{\mid Z\mid^2}$belongs to $\DD(q)$ iff
\begin{equation}
\alpha\in\Sigma(m)=\left\{\alpha\in\mathcal{A},\;\;m^2\alpha_0+\frac{\alpha_1}{2}-\frac{2\alpha_2}{m^2}=m^2\left(\frac{4\log2+3-4\gamma}{32}-\log m\right)\right\},
 \label{siguema}
\end{equation}
and for $m\geq 0$, the time-periodic solutions $\frac{e^{\pm imt}}{\mid Z\mid^4}$ belong to $\DD(q)$ iff
\begin{equation}
\alpha\in\Sigma(0)=\left\{(\alpha_0,\alpha_1,\alpha_2);\;\alpha_2=0,\;\alpha_1>0,\;\alpha_0+\frac{1}{2}\log\alpha_1<\frac{1}{4}-\frac{\log2}{2}-\gamma\right\}.
 \label{sigmazero}
\end{equation}
\end{Theorem}


{\it Proof of Theorem \ref{family}.}
With (\ref{vu}) we can check that we have
\begin{equation*}
q_{\lambda}(V_r(0,v_0,v_1,v_2)=U_r(0)+\lambda_0u_0-\frac{\gamma_1}{\mu_1-\mu_2}u_1-\frac{\gamma_2}{\mu_2-\mu_1}u_2
 \label{}
\end{equation*}
iff
\begin{equation}
\lambda_0=G_0(0)+\frac{\alpha_0}{32\pi^3},
 \label{zozo}
\end{equation}
\begin{equation}
\gamma_1=(\mu_1-\mu_2)\left[\frac{\alpha_0}{64\pi^3}\mu_1^2-\frac{\alpha_1}{16\pi^3}\mu_1-\frac{\alpha_2}{4\pi^3}-F_1(0)\right],
 \label{gagun}
\end{equation}
\begin{equation}
\gamma_2=(\mu_2-\mu_1)\left[\frac{\alpha_0}{64\pi^3}\mu_2^2-\frac{\alpha_1}{16\pi^3}\mu_2-\frac{\alpha_2}{4\pi^3}-F_2(0)\right].
 \label{gagde}
\end{equation}
Equation (\ref{zozo}) yields no constraint on $\alpha_j$ since $\lambda_0$ is an arbitrary real number. In opposite, (\ref{gagun}) and (\ref{gagde}) show that $\alpha=(\alpha_0,\alpha_1,\alpha_2)$ defines a linear form of the families $q_{\lambda}$, if and only if we can find $\mu_1,\mu_2<0$, $\mu_1\neq\mu_2$, $\gamma_1, \gamma_2>0$ solutions of these equations that we can write as :
\begin{equation}
\gamma_1=\frac{\mu_1-\mu_2}{16\pi^3}\mu_1^2G_{\alpha}(\mu_1),\;\;\gamma_2=\frac{\mu_2-\mu_1}{16\pi^3}\mu_2^2G_{\alpha}(\mu_2),
 \label{ekqc}
\end{equation}
where
\begin{equation*}
G_{\alpha}(\mu)=\frac{\alpha_0}{4}-\frac{\alpha_1}{\mu}-\frac{4\alpha_2}{\mu^2}+\frac{1}{8}\log(\mid\mu\mid)-\frac{\log2}{4}-\frac{3}{16}+\frac{\gamma}{4}.
 \label{}
\end{equation*}
We note that the conditions $\gamma_j>0$ in (\ref{ekqc}) are equivalent to the constraint
\begin{equation*}
\exists \mu_i,\mu_j;\;\; \mu_i<\mu_j<0,\;\;G_{\alpha}(\mu_i)<0<G_{\alpha}(\mu_j).
 \label{}
\end{equation*}
An elementary study of the function $G_{\alpha}$ shows that this case occurs iff
\begin{equation}
\exists \mu_*<0,\;\;G_{\alpha}(\mu_*)=0,\;\;G'_{\alpha}(\mu_*)>0.
 \label{zepb}
\end{equation}
Since $G_{\alpha}'(\mu)=\mu^{-3}(\frac{1}{8}\mu^2+\alpha_1\mu+8\alpha_2)$, we look for the $\alpha$ such that
\begin{equation}
\exists \mu_*<0,\;\;G_{\alpha}(\mu_*)=0,\;\;\frac{1}{8}\mu_*^2+\alpha_1\mu_*+8\alpha_2<0.
 \label{zepb}
\end{equation}
If $\mu_{\pm}:=4\left(-\alpha_1\pm\sqrt{\alpha_1^2-4\alpha_2}\right) $, an obvious equivalent condition is
\begin{equation*}
\alpha_1^2>4\alpha_2,\;\;\exists\mu_*\in \left]\mu_-,\mu_+\right[\cap]-\infty,0[,\;\;G_{\alpha}(\mu_*)=0. 
 \label{}
\end{equation*}
Therefore, taking account of the asymptotic behaviour of $G_{\alpha}(\mu)$ as $\mu\rightarrow 0^-$,  we have to determine the set of $\alpha$ such that $G_{\alpha}(\mu_-)<0$, and $G_{\alpha}(\mu_+)>0$ when $\mu_+<0$. The constraints (\ref{zin}), (\ref{zinb}) easily follow.

Now we prove that different $\alpha$ yield to different dynamics. We can see that $$u(t,Z)=V_r(t,Z)+v_0(t)\chi(Z)\log(\mid Z\mid)+v_1(t)\frac{\chi(Z)}{\mid Z\mid^2}+v_2(t)\frac{\chi(Z)}{\mid Z\mid^4}$$
is  solution of (\ref{equu}) iff
\begin{equation*}
\begin{split}
0=&\partial_t^2V_r-\Delta V_r+m^2V_r-\left(v_0\log(\mid Z\mid)+\frac{v_1}{\mid Z\mid^2}+\frac{v_2}{\mid Z\mid^4}\right)\Delta\chi\\
&-\left(2\frac{v_0}{\mid Z\mid^2}-4\frac{v_1}{\mid Z\mid^4}-4\frac{v_2}{\mid Z\mid^6}\right)Z.\nabla\chi+(\ddot{v}_0+m^2v_0)\chi\log(\mid Z\mid)\\
&+(\ddot{v}_1+m^2v_1-4v_0)\frac{\chi}{\mid Z\mid^2}+(\ddot{v}_2+m^2v_2+4v_1)\frac{\chi}{\mid Z\mid^4}.
\end{split}
 \label{}
\end{equation*}

When $u\in C^2(\RR_t;\HH_0)$, we have $V_r+v_0\chi\log(\mid Z\mid)\in C^2(\RR_t;H^2(\RR^6))$, $v_1$, $v_2\in C^2(\RR)$.
$u\in C^1(\RR_t;\HH_1)$ implies that $v_0\in C^1(\RR)$ and $V_r\in C^1(\RR_t;H^3(\RR^6))$. Finally $u\in C^0(\RR_t; \DD(q_{\lambda}))$ yields $V_r\in C^0(\RR_t;H^4(\RR^6))$. Now we consider a strong solution $u$ of which the initial data are  two spherically symmetric functions $f$, $g$ in $C^{\infty}_0(\RR^6\setminus\{0\})$, $(f,g)\neq(0,0)$. We know that there exists $T$ such that $v_2(T)\neq 0$. Taking account of the regularity of each terms in the previous equation, we get that
\begin{equation}
\begin{split}
0=&\partial_t^2V_r-\Delta V_r+m^2V_r-\left(v_0\log(\mid Z\mid)+\frac{v_1}{\mid Z\mid^2}+\frac{v_2}{\mid Z\mid^4}\right)\Delta\chi\\
&-\left(2\frac{v_0}{\mid Z\mid^2}-4\frac{v_1}{\mid Z\mid^4}-4\frac{v_2}{\mid Z\mid^6}\right)Z.\nabla\chi+(\ddot{v}_0+m^2v_0)\chi\log(\mid Z\mid),
\end{split}
 \label{ekv}
\end{equation}
\begin{equation}
0=\ddot{v}_1+m^2v_1-4v_0,
 \label{ekvun}
\end{equation}
\begin{equation}
0=\ddot{v}_2+m^2v_2+4v_1.
 \label{ekvde}
\end{equation}
We assume that $u$ is  solution associated with two linear forms with $(\alpha_0,\alpha_1,\alpha_2)$ and $(\alpha_0',\alpha_1',\alpha_2')$. Then we have
$$
(\alpha_0-\alpha_0')v_0(t)+(\alpha_1-\alpha_1')v_1(t)+(\alpha_2-\alpha_2')v_2(t)=0.
$$
If $\alpha_0\neq \alpha_0'$ we can express $v_0$ in terms of $v_1$ and $v_2$ in (\ref{ekvun}) and with (\ref{ekvde}) and the initial data $v_j(0)=\dot{v}_j(0)=0$ we obtain $v_1(t)=v_2(t)=0$ for all $t$, that is a contradiction with $v_2(T)\neq 0$. We deduce that $\alpha_0=\alpha_0'$. Now if $\alpha_1\neq\alpha_1'$, we express $v_1$ by $-\frac{\alpha_2-\alpha_2'}{\alpha_1-\alpha_1'}v_2$ in (\ref{ekvde}) and we obtain $v_2=0$ again, hence $\alpha_1=\alpha_1'$ and $(\alpha_2-\alpha_2')v_2=0$. Finally since $v_2(T)\neq 0$ we conclude that $\alpha_2=\alpha_2'$.\\

Finally to determine $\Sigma(m)$ we use (\ref{delk}) to get the components of the static solution $\mid Z\mid^{-2}K_2(m\mid Z\mid)$ : $v_2=\frac{2}{m^2}$, $v_1=-\frac{1}{2}$, $v_0=-m^2$, $V_r(0)=-m^2\log m+m^2 F(0)$ and the result follows from (\ref{F}). To characterize $\Sigma(0)$, we note that $V_r(0)=v_0=v_1=0$ and $v_2=e^{imt}$ for the time periodic solution $u(t,Z)=\mid Z\mid^{-4}e^{\pm imt}$. Hence $u(t,.)\in\DD(q)$ iff $\alpha_2=0$, and we conclude with (\ref{zin}) and (\ref{zinb}).

\fin

\section{Super-singular perturbations of the $1+1$D-Klein-Gordon equation}
In this section we investigate the Cauchy problem for some super-singular perturbations of the Klein-Gordon equation on the half line with a Bessel potentiel and a mass $m\geq 0$ :
\begin{equation}
\left\{
\begin{array}{c}
\partial_t^2\psi-\partial_z^2\psi+\frac{15}{4z^2}\psi+m^2\psi=0,\;\;t\in\RR,\;\;z>0,\\
\psi(0,z)=f(z),\;\;\partial_t\psi(0,z)=g(z)\;\;z>0.
\end{array}
\right.
 \label{zepbun}
\end{equation}
We recall some basic facts (see e.g. \cite{braneg} p. 532). The Bessel operator
\begin{equation}
P_2:=-\frac{d^2}{dz^2}+\frac{15}{4z^2}
 \label{pdeux}
\end{equation}
with domain $C^{\infty}_0(]0,\infty[)$ is essentially self-adjoint in $L^2(0,\infty)$ since $15/4\geq 3/2$ and its unique self-adjoint extension is the Friedrichs extension $\mathbf A_F$  of which the domain is 
\begin{equation*}
{\mathfrak d}_F:=\left\{\psi\in L^2(0,\infty);\;P_2\psi\in L^2\right\}=\left\{\psi\in L^2(0,\infty);\;P_2\psi,\;\psi',\,z^{-1}\psi\in L^2\right\}.
 \label{}
\end{equation*}
As a consequence, the Cauchy problem is well-posed for $f\in H^1_0(]0,\infty[)$, $g\in L^2(0,\infty)$ and the solution $\psi\in C^0(\RR_t;H^1_0(]0,\infty[)\cap C^1(\RR_t;L^2(0,\infty))$ is given by the standard formula
$$
\psi(t)=\cos\left(t\sqrt{{\mathbf A_F}+m^2}\right)f+\frac{\sin\left(t\sqrt{{\mathbf A_F}+m^2}\right)}{\sqrt{{\mathbf A_F}+m^2}}g.
$$
These solutions are called {\it ``Friedrichs solutions''} of (\ref{zepbun}) and they satisfy the conservation of the natural energy
$$
{\mathcal E}(\psi):=\int_0^{\infty}\mid\partial_t\psi(t,z)\mid^2+\mid\partial_z\psi(t,z)\mid^2+\left(m^2+\frac{15}{4z^2}\right)\mid\psi(t,z)\mid^2dz,
$$
and the Dirichlet condition at the origin :
$$
\psi(t,0)=0.
$$
We want to construct other solutions of (\ref{zepbun}) associated with other energies and other constraints at $z=0$. We could use the recent spectral results on the singular perturbations of the Bessel operators in \cite{kurasov-luger} but an easier way consists in using the link of $P_2$ and the Laplace operator in $\RR^6$,
$$
-\Delta_Z=z^{-\frac{5}{2}}\left(P_2-\frac{1}{z^2}\Delta_{S^5}\right)z^{\frac{5}{2}}.
$$
In this way, we can apply the results of the previous section. Then the super-singular perturbations of $\Delta_Z$ restricted to the spherically symmetric functions, yield to hypersingular perturbations of $P_2$ in the spaces of the trace of the radial distributions (see \cite{sickel} for an extensive study of these spaces).\\

Now we perform the suitable functional framework.
We introduce the differential operators
\begin{equation}
P_1:=\frac{d}{dz}-\frac{5}{2z},\;\;P_1^*:=-\frac{d}{dz}-\frac{5}{2z},
 \label{pinpin}
\end{equation}
and, for $1\leq k\leq 4$, we define the Hilbert spaces ${\bf h}^k$ as the closure of $C^{\infty}_0(]0,\infty[)$ for the following norms :
\begin{equation}
k=1,2,\;\;\Vert\psi\Vert_{{\bf h}^k}^2:=\Vert \psi\Vert_{L^2}^2+\Vert P_k\psi\Vert_{L^2}^2,
\;\;
\Vert\psi\Vert_{{\bf h}^{k+2}}^2:=\Vert\psi\Vert_{L^2}^2+\Vert P_kP_2\psi\Vert_{L^2}^2.
 \label{ashka}
\end{equation}
Given $\chi\in C^{\infty}_0(\RR)$ such that $\chi(z)=1$ in a neighborhood of $z=0$, we introduce the spaces
\begin{equation}
k=-1,0,\;\;\mathfrak{h}_k:=\left\{\psi(z)=\psi_r(z)+v_1\chi(z)z^{\frac{1}{2}}+v_2\chi(z)z^{-\frac{3}{2}},\;\;\psi_r\in{\bf h}^{k+2},\;v_j\in\CC\right\},
 \label{cheka}
\end{equation}
\begin{equation}
\mathfrak{h}_1:=\left\{\psi(z)=\psi_r(z) +v_0\chi(z)z^{\frac{5}{2}}\log z+v_1\chi(z)z^{\frac{1}{2}}+v_2\chi(z)z^{-\frac{3}{2}},\;\;\psi_r\in{\bf h}^{3},\;v_j\in\CC\right\},
 \label{chiun}
\end{equation}
\begin{equation}
\mathfrak{h}_2:=\left\{\psi(z)=\psi_r(z) +v_{-1}\chi(z)z^{\frac{5}{2}}+v_0\chi(z)z^{\frac{5}{2}}\log z+v_1\chi(z)z^{\frac{1}{2}}+v_2\chi(z)z^{-\frac{3}{2}},\;\psi_r\in{\bf h}^{4},\;v_j\in\CC\right\},
 \label{cjoui}
\end{equation}
and if $X$ a space of distributions on $\RR^6_Z$, we introduce the subspace $RX$ of the spherically symmetric distributions of $X$ :
\begin{equation*}
RX:=\left\{u\in X;\;\;Z_i\partial_{Z_j}u-Z_j\partial_{Z_i}u=0,\;\;1\leq i<j\leq 6\right\}.
 \label{}
\end{equation*}
Given $u\in L^2(\RR^6_Z)$ we associate $\psi_u$ defined on $]0,\infty[_z$ by
\begin{equation}
\psi_u(\mid Z\mid):=\mid Z\mid^{\frac{5}{2}}u(Z).
 \label{fiu}
\end{equation}

\begin{Lemma}
 \label{lemrad}
Given $\psi\in L^2(0,\infty)$, $\psi$ belongs to ${\bf h}^k$ if and only if $u_{\psi}(Z):=\mid Z\mid^{-\frac{5}{2}}\psi(\mid Z\mid)$ belongs to $H^k(\RR^6_Z)$ and $u_{\psi}(0)=0$ for $k=4$. As a consequence, we have
\begin{equation*}
{\bf h}^4\subset {\bf h}^3\subset {\bf h}^2\subset {\bf h}^1,
 \label{}
\end{equation*}
\begin{equation}
\psi \in {\bf h}^1,\;\mid\psi(z)\mid \leq Cz^{\frac{1}{2}},
\;\;
\psi \in {\bf h}^2,\;\mid\psi(z)\mid \leq Cz^{\frac{3}{2}},
 \label{dikk}
\end{equation}
\begin{equation}
\psi \in {\bf h}^3,\;\mid\psi(z)\mid \leq Cz^{\frac{5}{2}}\sqrt{\mid\log z\mid},
\;\;
\psi \in {\bf h}^4,\;\lim_{z\rightarrow0^+}z^{-\frac{5}{2}}\psi(z)=0,
 \label{dikkk}
\end{equation}
\begin{equation}
-1\leq k\leq 2,\;\;{\mathfrak h}_k=\left\{\psi_u;\;\;u\in R\HH_k\right\}.
 \label{hhh}
\end{equation}
The coefficients $v_j$ do not depend on the choice of the function $\chi$ and $v_{-1}=V_r(0)$ when $\psi\in{\mathfrak h}_2$ and $u_{\psi}=V_r+v_0\chi\log\mid Z\mid+v_1\chi\mid Z\mid^{-2}+v_2\chi\mid Z\mid^{-4}$. The spaces $\mathfrak{h}_k$ are Hilbert spaces for the norms
\begin{equation}
\Vert\psi\Vert_{{\mathfrak h}_k}^2:=\Vert\psi_r\Vert_{{\bf h}^{k+2}}^2+\sum_j\mid v_j\mid^2.
 \label{normah}
\end{equation}
\end{Lemma}

{\it Proof of Lemma \ref{lemrad}.}
We remark that for $u\in RC^{\infty}_0(\RR^6\setminus\{0\})$ we have
\begin{equation}
\int_{\RR^6}\mid u(Z)\mid^2dZ=\pi^3\int_0^{\infty}\mid\psi_u(z)\mid^2dz,\;\;
\int_{\RR^6}\mid \nabla_Zu(Z)\mid^2dZ=\pi^3\int_0^{\infty}\mid\psi'_u(z)-\frac{5}{2z}\psi_u(z)\mid^2dz,
 \label{fufi}
\end{equation}
$$
\int_{\RR^6}\mid \Delta_Zu(Z)\mid^2dZ=\pi^3\int_0^{\infty}\mid\psi''_u(z)-\frac{15}{4z^2}\psi_u(z)\mid^2dz.
$$
We deduce that $u\mapsto\pi^{-\frac{3}{2}}\psi_u$ is an isometry from $ RC^{\infty}_0(\RR^6\setminus\{0\})$ endowed with a suitable $H^k(\RR^6)$-norm, into  $ RC^{\infty}_0(]0,\infty[)$ endowed with the ${\bf h}_k$-norm. Since $ RC^{\infty}_0(\RR^6\setminus\{0\})$ is dense in  $H^m(\RR^6)$ for $m\leq 3$, we conclude that
\begin{equation}
k=-1,0,1,\;\;{\bf h}_{k+2}=\left\{\psi_u;\;\;u\in RH^{k+2}(\RR^6)\right\},\;\;{\mathfrak h}_k=\left\{\psi_u;\;\;u\in R\HH_k\right\}
 \label{chouchou}
\end{equation}
and  (\ref{normah}) defines a norm $\Vert\psi_u\Vert_{{\mathfrak h}_k}\sim\Vert u\Vert_{\HH_k}$, for which  ${\mathfrak h}_k$ is a Hilbert space.

On the other hand, the Sobolev embedding $H^4(\RR^6)\subset C^0(\RR^6)$ implies that the closure of $ RC^{\infty}_0(\RR^6\setminus\{0\})$ in $RH^4(\RR^6)$ is the set of functions $u\in RH^4$ that are zero at $Z=0$, and $RH^4(\RR^6)=\overline{ RC^{\infty}_0(\RR^6\setminus\{0\})}\oplus\CC\chi(\mid Z\mid)$. We conclude that $\lim_{z\rightarrow0^+}z^{-\frac{5}{2}}\psi(z)=0$ when $\psi\in {\bf h}_4$ and 
$$
{\bf h}_4=\left\{\psi_u;\;\;u\in RH^4(\RR^6),\;u(0)=0\right\},\;{\bf h}_4\oplus\CC\chi(z)z^{\frac{5}{2}}=\left\{\psi_u;\;u\in RH^4(\RR^6)\right\},\;
{\mathfrak h}_2=\left\{\psi_u;\;\;u\in R\HH_2\right\}.
$$
The decay near the origin (\ref{dikk}),  (\ref{dikkk})  for $k=1,2,3$  are consequences of theorems 13 and 14 of \cite{sickel}. To achieve the proof of the lemma, we remark that
$\chi(z)z^{\frac{1}{2}}\notin {\bf h}_1$, $\chi(z)z^{\frac{5}{2}}\log z\in {\bf h}_2\setminus{\bf h}_3$, $\chi(z)z^{\frac{5}{2}}\in {\bf h}_3\setminus{\bf h}_4$. Then the coefficients $v_j$ only depend on $\psi$ and since $V_r(Z)=\mid Z\mid^{-\frac{5}{2}}\psi_r(\mid Z\mid)=+v_{-1}\chi(\mid Z\mid)$, we have $v_{-1}=V_r(0)$. Finally since $\Vert u\Vert_{H^4}\sim\Vert\psi_r\Vert_{{\mathbf h}_4}+\mid v_{-1}\mid$ for $u\in RH^4(\RR^6)$, we have $\Vert u\Vert_{\HH_2}\sim\Vert\psi_u\Vert_{{\mathfrak h}_2}$ and it is clear that (\ref{normah}) defines a norm for which  ${\mathfrak h}_2$ is a Hilbert space.

\fin

We now introduce the ``boundary conditions''. Given $\alpha=(\alpha_0,\alpha_1,\alpha_2)\in\RR^3$, we consider the Hilbert subspace
\begin{equation}
\mathfrak{d}_{\alpha}:=\left\{\psi\in\mathfrak{h}_2;\;\;v_{-1}+\alpha_0v_0+\alpha_1v_1+\alpha_2v_2=0\right\},
 \label{dalfa}
\end{equation}
and we denote $\mathbf A_{\alpha}$ the differential operator $P_2$ endowed with  $\mathfrak{d}_{\alpha}$ as domain.
The existence of super-singular perturbations of the Bessel operator $P_2$ is stated by the following :

\begin{Proposition}
 \label{propbessel}
  For all $\alpha=(\alpha_0,\alpha_1,\alpha_2)\in\RR^3$ satisfying the constraints (\ref{zin}) and (\ref{zinb}), there exists a hermitian product on ${\mathfrak h}_0$, equivalent to the initial $\Vert.\Vert_{{\mathfrak h}_0}$-scalar product, for which  $\mathbf A_{\alpha}$  is a semi-bounded from below, self-adjoint operator on ${\mathfrak h}_0$. Its essential spectrum is $[0,\infty[$.  Its point spectrum is a set of $0$, $1$, $2$ or $3$ non positive eigenvalues $-\lambda_j^2$, associated with eigenfunctions $\psi_j(z)=\sqrt{z}K_2(\lambda_jz)$ if $\lambda_j>0$, $\psi_j(z)=z^{-\frac{3}{2}}$ if $\lambda_j=0$. Moreover $\lambda_j^2>0$ are the roots of the equation :
\begin{equation}
\log x+2\alpha_0+\frac{8\alpha_1}{x}-\frac{32\alpha_2}{x^2}=0,
 \label{funclam}
\end{equation}
and $0$ is eigenvalue iff $\alpha$ belongs to $\Sigma(0)$ defined by (\ref{sigmazero}). In particular, the point spectrum is empty for all $\alpha$ such that
\begin{equation}
\alpha_2<0,\;\;-\log2<\alpha_0+\frac{\alpha_1}{\alpha_1+\sqrt{\alpha_1^2-4\alpha_2}}-\frac{\alpha_2}{\left(\alpha_1+\sqrt{\alpha_1^2-4\alpha_2}\right)^2}+\frac{1}{2}\log\left(\alpha_1+\sqrt{\alpha_1^2-4\alpha_2}\right)<\frac{3}{4}-\gamma,
 \label{spev}
\end{equation}
and $0$ is the unique eigenvalue when
\begin{equation}
\alpha_2=0<\alpha_1,\;-\frac{1}{2}-\frac{3}{2}\log2<\alpha_0+\frac{1}{2}\log\alpha_1<\frac{1}{4}-\frac{1}{2}\log2-\gamma
 \label{spevo}
\end{equation}
\end{Proposition}

{\it Proof of Proposition \ref{propbessel}.}
The previous lemma assures that the map $\psi\mapsto u_{\psi}$ defined by
\begin{equation}
\psi(z)=\psi_r(z)+v_1\chi(z)z^{\frac{1}{2}}+v_2\chi(z)z^{-\frac{3}{2}}\longmapsto
u_{\psi}(Z)=\frac{\psi_r(\mid Z\mid)}{\mid Z\mid^{\frac{5}{2}}}+v_1\frac{\chi(\mid Z\mid )}{\mid Z\mid^2}+v_2\frac{\chi(\mid Z\mid )}{\mid Z\mid^4}
 \label{ufi}
\end{equation}
is an isometry from ${\mathfrak h}_0$ onto $R\HH_0$, where $\HH_0$ is the space (\ref{ho}) endowed with the equivalent norm
$$
\pi^{-\frac{3}{2}}\left(\parallel v_r\parallel_{L^2(\RR^6)}^2+\parallel \Delta v_r\parallel_{L^2(\RR^6)}^2+\sum_{j=1}^2\mid v_j\mid^2\right)^{\frac{1}{2}}.
$$
Moreover we have for any $\psi\in{\bf h}^1$
\begin{equation*}
u_{P_2\psi}=-\Delta u_{\psi}-4\pi^3v_2\delta_0(Z).
 \label{}
\end{equation*}
Now we consider $\mu_1,\mu_2<0$, $\mu_1\neq\mu_2$, $\lambda_0\in\RR$, $\gamma_1,\gamma_2>0$ associated with $\alpha$ by the Theorem \ref{family}, and we endow $\HH_0$ with the norm $\Vert.\Vert_0$ given by (\ref{normho}) for which $\AA_{\lambda}$ defined by (\ref{domal}) and (\ref{agagaga}) is semi-bounded from below, self-adjoint. We remark that $$
\left(Z_i\partial_j-Z_j\partial_i\right)\left(-\Delta_Z+L(u)\delta_0\right)=-\Delta_Z=\left(-\Delta_Z+L(u)\delta_0\right) \left(Z_i\partial_j-Z_j\partial_i\right),
$$
hence the restriction of $\AA_{\lambda}$ to $R\HH_0$ with the domain $RDom(\AA_{\lambda})$ is a densely defined self-adjoint operator that we denote  $R\AA_{\lambda}$.
Since
$${\mathfrak d}_{\alpha}=\left\{\psi_u;\;\;u\in RDom(\AA_{\lambda})\right\},
\;\;
\AA_{\lambda}u_{\psi}=u_{P_2\psi}
$$
we conclude that if ${\mathfrak h}_0$ is endowed with the equivalent norm
\begin{equation}
\Vert\psi\Vert_0:=\Vert u_{\psi}\Vert_0,
 \label{normoad}
\end{equation}
where $\Vert u_{\psi}\Vert_0 $ is defined by (\ref{normho}), then ${\mathbf A_{\alpha}}$ is unitarily equivalent to $R\AA_{\lambda}$. Therefore it is semi-bounded from below, and self-adjoint on ${\mathfrak h}_0$.
We introduce the operator $\mathbf{A}_0$ defined as the differential operator $P_2$ provided with
\begin{equation}
{\mathfrak d}_0:=\left\{\psi_u;\;u\in R\HH_2,\;u_0=0\right\}=\left\{\psi_u;\;u=U_r+u_1\varphi_1(Z)+u_2\varphi_2(Z),\;U_r\in RH^4(\RR^6_Z),\;u_j\in\CC\right\}.
 \label{dododo}
\end{equation}
Then $\mathbf{A}_0$ is unitarily equivalent to $R\AA_0$ where $\AA_0$ is given by (\ref{azero}). Since the essential spectrum of the Laplacien considered as an operator on $RH^2(\RR^6)$ endowed with its natural domain $RH^4(\RR^6)$ is $[0,\infty[$, and $\left(\mathbf{A}_0+i\right)^{-1}-\left(\mathbf{A}_{\alpha}+i\right)^{-1}$ is finite rank, we conclude by the Weyl theorem that $\sigma_{ess}(\mathbf{A}_{\alpha})=[0,\infty[$.\\

Now given $\lambda>0$, the solutions of $P_2\psi=\lambda^2\psi$ are given by
$\psi(z)=A\sqrt{z}J_2(\lambda z)+B\sqrt{z}Y_2(\lambda z)$. Since $\psi(z)\sim -\sqrt{\frac{2}{\pi}}\left[A\cos(z-\frac{\pi}{4})+B\sin(z-\frac{\pi}{4})\right]$, $\psi$ does not belong to ${\mathfrak h}_0$ when $(A,B)\neq (0,0)$. We conclude that the eigenvalues of $P_2$ are non positive. On the other hand, the solutions of $P_2\psi=-\lambda^2\psi$ are given by
$\psi(z)=A\sqrt{z}I_2(\lambda z)+B\sqrt{z}K_2(\lambda z)$. Since $I_2(z)\sim \frac{1}{\sqrt{2\pi z}}e^z$ as $z\rightarrow\infty$, and taking account of (\ref{delk}), the eigenfunction in ${\mathfrak h}_0$ is
$$
\psi(z)=\sqrt{z}K_2(\lambda z)=\lambda^4z^{\frac{9}{2}}G(\lambda^2z^2)\log(\lambda z)+\lambda^2z^{\frac{5}{2}}F(\lambda^2z^2)-\left(\frac{\lambda^2}{8}\log\lambda\right)z^{\frac{5}{2}}
-\frac{\lambda^2}{8}z^{\frac{5}{2}}\log z-\frac{1}{2}z^{\frac{1}{2}}+\frac{2}{\lambda^2}z^{-\frac{3}{2}}.
$$
Then $v_{-1}=-\frac{\lambda^2}{8}\log\lambda$, $v_0=-\frac{\lambda^2}{8}$, $v_1=\frac{1}{2}$, $v_2=\frac{2}{\lambda^2}$ satisfy $v_{-1}+\alpha_0v_0+\alpha_1v_1+\alpha_2v_2=0$ iff $\lambda^2$ is a stricly positive solution of (\ref{funclam}). To determine the number of these roots, we study the fonction $h(x):=\log x+2\alpha_0+\frac{8\alpha_1}{x}-\frac{32\alpha_2}{x^2}$. When $\alpha_2<0$, or when $\alpha_2=0$ and $\alpha_1>0$, $h$ is decreasing from $+\infty$ to $\inf h=2\left(\alpha_0+\frac{\alpha_1}{\alpha_1+\sqrt{\alpha_1^2-4\alpha_2}}-\frac{\alpha_2}{(\alpha_1+\sqrt{\alpha_1^2-4\alpha_2})^2}+\frac{1}{2}\log(\alpha_1+\sqrt{\alpha_1^2-4\alpha_2})+log2\right)$ when $x\in]0,4(\alpha_1+\sqrt{\alpha_1^2-4\alpha_2})]$, and from $\inf h$ to $+\infty$ for $x\in[4(\alpha_1+\sqrt{\alpha_1^2-4\alpha_2}),\infty[$. We deduce that there exists $0$, $1$ or $2$ roots according to $\inf h>0$, $\inf h=0$, $\inf h<0$. Then (\ref{spev}) and (\ref{spevo}) follow from (\ref{zin}) and (\ref{sigmazero}). Finally when $0<4\alpha_2<\alpha_1^2$ and $0<\alpha_1$, $h$ is increasing from $-\infty$ to $2\left(\alpha_0+\frac{\alpha_1}{\alpha_1-\sqrt{\alpha_1^2-4\alpha_2}}-\frac{\alpha_2}{(\alpha_1-\sqrt{\alpha_1^2-4\alpha_2})^2}+\frac{1}{2}\log(\alpha_1-\sqrt{\alpha_1^2-4\alpha_2})+log2\right)$ when $x\in]0,4(\alpha_1-\sqrt{\alpha_1^2-4\alpha_2})]$, decreasing for $x\in]4(\alpha_1-\sqrt{\alpha_1^2-4\alpha_2}), 4(\alpha_1+\sqrt{\alpha_1^2+4\alpha_2})]$, and increasing to $+\infty$ for $x>4(\alpha_1+\sqrt{\alpha_1^2+4\alpha_2})$. We conclude that in this case there exists $1$, $2$ or $3$ strictly negative eigenvalues.

\fin

Now we consider the Cauchy problem (\ref{zepbun}). We look for the weak solutions with the {\it Anstatz}
\begin{equation}
\psi(t,z)=\psi_r(t,z) +v_0(t)\chi(z)z^{\frac{5}{2}}\log z+v_1(t)\chi(z)z^{\frac{1}{2}}+\phi_2(t)z^{-\frac{3}{2}},
 \label{ansatzpsi}
\end{equation}
$$
v_1,\;\phi_2\in C^2(\RR),\;\psi_r(t,z) +v_0(t)\chi(z)z^{\frac{5}{2}}\log z\in C^2(\RR_t;{\bf h}^{1})\cap C^1(\RR_t;{\bf h}^{2}), \;v_0\in C^0(\RR),\;\psi_r\in C^0(\RR_t;{\bf h}^{3}),
$$
and we want to construct the strong solutions that satisfy
$$
\psi_r(t,z)=\psi_R(t,z) +v_{-1}(t)\chi(z)z^{\frac{5}{2}}, \;v_{-1}\in C^0(\RR),\;\psi_R\in C^0(\RR_t;{\bf h}^{4}),
$$
and the boundary condition
$$
v_{-1}(t)+\alpha_0v_0(t)+\alpha_1v_1(t)+\alpha_2\phi_2(t)=0,\;\;t\in\RR.
$$

We can state the main result of this part :


\begin{Theorem}
 \label{theobess}
 Let $\alpha=(\alpha_0,\alpha_1,\alpha_2)$ be in $\RR^3$ satisfying the constraints (\ref{zin}) and (\ref{zinb}). Then for all $m\geq 0$ and any $f\in{\mathfrak h}_1$, 
$g\in{\mathfrak h}_0$, the Cauchy problem (\ref{zepbun}) has a unique solution
\begin{equation}
\psi_{\alpha}\in C^0\left(\RR_t;{\mathfrak h}_1\right)\cap C^1\left(\RR_t;{\mathfrak h}_0\right) \cap C^2\left(\RR_t;{\mathfrak h}_{-1}\right)\cap{\mathcal D}'\left(\RR_t;{\mathfrak d}_{\alpha}\right),
 \label{reglagla}
\end{equation}
and if $f,g\in C^{\infty}_0(]0,\infty[)$, $(f,g)\neq (0,0)$, then $v_2\neq 0$ hence $\psi_{\alpha}$ is not the Friedrichs solution and $\psi_{\alpha}\neq\psi_{\alpha'}$ if $\alpha\neq \alpha'$. Moreover there exists $C,K>0$ independent of $m\geq 0$ such that
\begin{equation}
\Vert\partial_t\psi_{\alpha}(t)\Vert_{{\mathfrak h}_0}+\Vert\psi_{\alpha}(t)\Vert_{{\mathfrak h}_1}+m\Vert\psi_{\alpha}(t)\Vert_{{\mathfrak h}_0}\leq C\left(\Vert g\Vert_{{\mathfrak h}_0}+\Vert f\Vert_{{\mathfrak h}_1}+m\Vert f\Vert_{{\mathfrak h}_0}\right)e^{(K-m^2)_+\mid t\mid},
 \label{contipsi}
\end{equation}
and for all $\Theta\in C^{\infty}_0(\RR_t)$ we have :
\begin{equation}
\Vert\int\Theta(t)\psi_{\alpha}(t)dt\Vert_{\mathfrak{h}_2}\leq C\left(\Vert g\Vert_{{\mathfrak h}_0}+\Vert f\Vert_{{\mathfrak h}_1}+m\Vert f\Vert_{{\mathfrak h}_0}\right)\int\ \left(\mid\Theta(t)\mid+\mid\Theta''(t)\mid\right)e^{(K-m^2)_+\mid t\mid}dt.
 \label{contipsid}
\end{equation}

There exists a conserved energy, i.e. a non trivial, continuous quadratic form $\mathcal{E}_{\alpha}$ defined on ${\mathfrak h}_1\oplus{\mathfrak h}_0$, that satisfies :
\begin{equation}
\forall t\in\RR,\;\;\mathcal{E}_{\alpha}\left(\psi_{\alpha}(t),\partial_t\psi_{\alpha}(t)\right)=\mathcal{E}_{\alpha}(f,g).
 \label{enrpsi}
\end{equation}
This energy is not positive definite in general but $\mathcal{E}_{\alpha}$  is equivalent to $\Vert f\Vert^2_{\mathfrak h_1}+\Vert g\Vert^2_{\mathfrak h_0}$  on $C^{\infty}_0(]0,\infty[)\oplus C^{\infty}_0(]0,\infty[)$ and given for $f,g\in C^{\infty}_0(]0,\infty[)$ by
\begin{equation}
\begin{split}
\mathcal{E}_{\alpha}(f,g)=&\Vert P_1P_2f\Vert^2_{L^2}-(\mu_1+\mu_2) \Vert P_2f\Vert^2_{L^2}+\mu_1\mu_2 \Vert P_1f\Vert^2_{L^2}\\
&+m^2\left(\Vert P_2f\Vert^2_{L^2}-(\mu_1+\mu_2) \Vert P_1f\Vert^2_{L^2}+\mu_1\mu_2 \Vert f\Vert^2_{L^2}\right)\\
&+\Vert P_2g\Vert^2_{L^2}-(\mu_1+\mu_2) \Vert P_1g\Vert^2_{L^2}+\mu_1\mu_2 \Vert g\Vert^2_{L^2}
\end{split}
 \label{enerex}
\end{equation}
for some $\mu_1<\mu_2<0$.
When $\alpha$ satisfies (\ref{spev}) or (\ref{spevo}), $\mathcal{E}_{\alpha}$ is positive  on ${\mathfrak h}_1\oplus{\mathfrak h}_0$.\\

The propagation is causal, i.e.
\begin{equation}
supp(\psi_{\alpha}(t,.))\subset\{z;\;\mid z\mid\leq\mid t\mid\}+\left[supp(f)\cup supp(g)\right].
 \label{causepsi}
\end{equation}

When $f\in {\mathfrak d}_{\alpha}$, $g\in {\mathfrak h}_1$ then $\psi_{\alpha}$ satisfies :
\begin{equation}
\psi_{\alpha}\in C^0\left(\RR_t;{\mathfrak d}_{\alpha}\right)\cap C^1\left(\RR_t;{\mathfrak h}_1\right)\cap C^2\left(\RR_t;{\mathfrak h}_0\right),
 \label{regloupi}
\end{equation}
\begin{equation}
\Vert\partial_t\psi_{\alpha}(t)\Vert_{{\mathfrak h}_1}+\Vert\psi_{\alpha}(t)\Vert_{{\mathfrak h}_2}+m\Vert\psi_{\alpha}(t)\Vert_{{\mathfrak h}_1}\leq C\left(\Vert g\Vert_{{\mathfrak h}_1}+\Vert f\Vert_{{\mathfrak h}_2}+m\Vert f\Vert_{{\mathfrak h}_1}\right)e^{(K-m^2)_+\mid t\mid}.
 \label{contipside}
\end{equation}

\end{Theorem}

{\it Proof of Theorem \ref{theobess}.}
We consider $\mu_1,\mu_2<0$, $\mu_1\neq\mu_2$ and $\lambda\in\RR^3$ that are  associated with $\alpha$ by the Theorem \ref{family}. We introduce $u_{\lambda}(t,Z):=\mid Z\mid^{\frac{5}{2}}\psi_{\alpha}(\mid Z\mid)$. Then Lemma \ref{lemrad} assures that $u_{\lambda}\in C^m(\RR_t;\HH_k)$ iff $\psi_{\alpha}\in C^m(\RR_t;\mathfrak{h}_k)$ and 
$u_{\lambda}\in{\mathcal D}'(\RR_t;\DD(q_{\lambda})$ iff $\psi_{\alpha}\in{\mathcal D}'(\RR_t;{\mathfrak d}_{\alpha})$. Moreover since ${\mathbf A_{\alpha}}$ is unitarily equivalent to $R\AA_{\lambda}$, we can see that $\psi_{\alpha}$ is the wanted solution iff $u_{\lambda}$ is the solution of (\ref{regul}) and (\ref{equu}) with the corresponding  initial data. Therefore Theorem \ref{theoun} gives the existence, the uniqueness, the estimates of the solution of the Cauchy problem  (\ref{zepbun}), and the finite velocity result (\ref{causepsi}). Moreover when $f,g\in C^{\infty}_0(]0,\infty[)$, $(f,g)\neq (0,0)$, $\psi_{\alpha}$ is not a Friedrichs solution since the dynamics for $u_{\lambda}$ is not trivial according Theorem \ref{theoun}, and  Theorem \ref{family} implies that different $\alpha$ yield to different solutions. Finally the energy is given for the strong solutions  by
\begin{equation}
{\mathcal E}_{\alpha}(\psi_{\alpha},\partial_t\psi_{\alpha}):=\left<{\bf A}_{\alpha}\psi_{\alpha};\psi_{\alpha}\right>_{0}+
m^2\left\Vert\psi_{\alpha}\right\Vert_{0}^2+\left\Vert\partial_t\psi_{\alpha}\right\Vert_{0}^2=\pi^3 {\mathcal E}_{\lambda}(u_{\lambda},\partial_tu_{\lambda}),
 \label{pschit}
\end{equation}
where the norm  $\Vert.\Vert_0$ is defined by (\ref{normoad}), and this energy is positive when $\alpha$ satisfies (\ref{spev}) or (\ref{spevo}) since the spectrum of ${\bf A}_{\alpha}$ is $[0,\infty[$ in this case by the Proposition \ref{propbessel}. At last, the expression (\ref{enerex}) is obtained by a direct computation by using the facts that $P_1^*P_1=P_2$ and for $u_{\lambda}\in C^{\infty}_0(\RR^6\setminus\{0\})$ we have :
$$
\Vert u_{\lambda}\Vert_{H^2}^2=\pi^3\left<(P_2-\mu_1)\psi_{\alpha};(P_2-\mu_2)\psi_{\alpha}\right>_{L^2(0,\infty)},
$$
$$
\Vert \nabla_Zu_{\lambda}\Vert_{H^2}^2=\pi^3\left<P_1(P_2-\mu_1)\psi_{\alpha};P_1(P_2-\mu_2)\psi_{\alpha}\right>_{L^2(0,\infty)}.
$$

\fin

We end this part by some remarks. Firstly, we note that when $\alpha$ satisfies (\ref{spev}) or (\ref{spevo}), the operator  ${\mathbf A_{\alpha}}$ is a positive self-adjoint operator in $({\mathfrak h}_0, \Vert.\Vert_0)$. Then, in this case, the solution is just given by the spectral functional calculus :
$$
\psi_{\alpha}(t,.)=\cos\left(t\sqrt{{\mathbf A_{\alpha}}+m^2}\right)f+\frac{\sin\left(t\sqrt{{\mathbf A_{\alpha}}+m^2}\right)}{\sqrt{{\mathbf A_{\alpha}}+m^2}}g,
$$
and we can solve the Cauchy problem in the scale of the Hilbert spaces associated with the powers of ${\mathbf A_{\alpha}}$. More precisely, when  $m>0$ or when $\alpha$ satisfies (\ref{spev}), the Cauchy problem is well-posed for $f\in\left[Dom\left(\left({\mathbf A_{\alpha}}+m^2\right)^{\frac{s+1}{2}}\right)\right]$, $g\in\left[Dom\left(\left({\mathbf A_{\alpha}}+m^2\right)^{\frac{s}{2}}\right)\right]$ where $[Dom(B)]$ denotes the completion of $Dom(B)$ for the norm $\Vert B.\Vert_0$.   Secondly, when $\alpha$ satisfies (\ref{spevo}), the kernel of ${\mathbf A_{\alpha}}$ is $\CC z^{-\frac{3}{2}}$  and the time-periodic solutions $e^{\pm imt}z^{-\frac{3}{2}}$ belong to $C^0\left(\RR_t;{\mathfrak d}_{\alpha}\right)$. We can express $\psi_{\alpha}$ in term of the graviton part supported by $z^{-\frac{3}{2}}$:
\begin{equation}
\psi_{\alpha}(t,z)=\psi_{\alpha}^0(t) z^{-\frac{3}{2}}+\psi_{\alpha}^{\perp}(t,z),\;\;\left<\psi_{\alpha}^{\perp}(t,.);z^{-\frac{3}{2}}\right>_0=0,
 \label{gravexpand}
\end{equation}
where the amplitude of the graviton is given by
\begin{equation}
\psi_{\alpha}^0(t)=\Vert z^{-\frac{3}{2}}\Vert_0^{-2}\left(\cos(mt)<f;z^{-\frac{3}{2}}>_0+\frac{\sin(mt)}{m}<g;z^{-\frac{3}{2}}>_0\right),
 \label{graviton}
\end{equation}
and we have to replace $\frac{\sin(mt)}{m}$ by $t$ when $m=0$. Finally, if we could establish the absence of singular continuous spectrum of ${\mathbf A_{\alpha}}$, then $\psi_{\alpha}^{\perp}(t,.)$ would tend weakly to zero as $\mid t\mid\rightarrow\infty$, hence its component on $z^{-\frac{3}{2}}$ would become negligible for large time. An interesting consequence would be
\begin{equation}
 \phi_2(t)-\psi_{\alpha}^0(t)\rightarrow 0,\;\;\mid t\mid\rightarrow\infty,
 \label{scatgr}
\end{equation}
{\it i.e.} the more singular part in the expansion (\ref{ansatzpsi}) would be asymptotically given by the graviton.

\section{New dynamics in $AdS^5$}
In this section we construct new unitary dynamics for the gravitational waves in the Anti-de Sitter universe. We consider the Cauchy problem
\begin{equation}
\left(\partial_t^2-\Delta_{\mathbf{x}}-\partial_z^2+\frac{15}{4z^2}\right)\Phi=0,\;\;t\in\RR,\;\;{\mathbf x}\in\RR^3,\;\;z\in]0,\infty[,
  \label{eqad}
\end{equation}
\begin{equation}
\Phi(0,{\mathbf x},z)=\Phi_0({\mathbf x},z),\;\;\partial_t\Phi(0,{\mathbf x},z)=\Phi_1({\mathbf x},z).
 \label{ciad}
\end{equation}
We look for the solutions that have an expansion of the following form
\begin{equation}
\Phi(t,{\mathbf x},z)=\Phi_r(t,{\mathbf x},z)z^{\frac{5}{2}}+\phi_{-1}(t,{\mathbf x})\chi(z)z^{\frac{5}{2}}
+\phi_0(t,{\mathbf x})\chi(z)z^{\frac{5}{2}}\log z+\phi_{1}(t,{\mathbf x})\chi(z)z^{\frac{1}{2}}+\phi_{2}(t,{\mathbf x})z^{-\frac{3}{2}}
 \label{decoad}
\end{equation}
where $\chi\in C^{\infty}_0(\RR)$, $\chi(z)=1$ in a neighborhood of $0$ and
\begin{equation}
\Phi_r(t,{\mathbf x},0)=0.
 \label{cfri}
\end{equation}
The term $\phi_{2}(t,{\mathbf x})z^{-\frac{3}{2}}$ is the part of the wave in the sector of the massless graviton. The behaviour of the field on the boundary of the universe is assumed to be for some $(\alpha_0,\alpha_1,\alpha_2)\in\RR^3$ :
\begin{equation}
\phi_{-1}(t,{\mathbf x})
+\alpha_0\phi_0(t,{\mathbf x})+\alpha_1\phi_{1}(t,{\mathbf x})+\alpha_2\phi_{2}(t,{\mathbf x})=0,\;\;t\in\RR,\;\;{\mathbf x}\in\RR^3.
 \label{constraint}
\end{equation}
We introduce the following Hilbert spaces endowed with the natural norms (${\mathfrak h}_0$ being provided with the norm (\ref{normoad})) :
\begin{equation}
\begin{split}
{\mathfrak H}_0:=&L^2\left(\RR^3_{\mathbf x};{\mathfrak h}_0\right)\\
=&\left\{\Phi ({\mathbf x} ,z)=\phi_r ({\mathbf x} ,z)+\phi_1 ({\mathbf x})\chi(z)z^{\frac{1}{2}}+\phi_2 ({\mathbf x})z^{-\frac{3}{2}},\;\;\phi_r\in L^2(\RR^3_{\mathbf x};{\bf h}^{2}),\;\phi_j\in L^2(\RR^3_{\mathbf x})\right\},
\end{split}
 \label{HUIO}
\end{equation}
\begin{equation}
\begin{split}
{\mathfrak H}_1:=&\left\{\Phi\in L^2\left(\RR^3_{\mathbf x};{\mathfrak h}_1\right);\;\nabla_{\mathbf x}\Phi\in{\mathfrak H}_0\right\}\\
=&\left\{\Phi ({\mathbf x} ,z)=\phi_r ({\mathbf x} ,z)+\phi_0({\mathbf x})\chi(z)z^{\frac{5}{2}}\log z+\phi_1 ({\mathbf x})\chi(z)z^{\frac{1}{2}}+\phi_2 ({\mathbf x})z^{-\frac{3}{2}},\right.\\
&\left.\phi_r\in L^2(\RR^3_{\mathbf x};{\bf h}^{3}),\;\;\phi_0\in L^2(\RR^3_{\mathbf x}),\;\phi_1,\phi_2\in H^1(\RR^3_{\mathbf x}),\;\nabla_{\mathbf x}\left(\phi_r+\phi_0\chi z^{\frac{5}{2}}\log z\right)\in L^2(\RR^3_{\mathbf x};{\bf h}^{2})\right\},
\end{split}
 \label{HUIN}
\end{equation}
\begin{equation}
{\mathfrak H}_2:=\left\{\Phi\in L^2\left(\RR^3_{\mathbf x};{\mathfrak h}_2\right);\;\nabla_{\mathbf x}\Phi\in{\mathfrak H}_1\right\}.
 \label{HDE}
\end{equation}

In particular, $\Phi\in {\mathfrak H}_2$ iff
\begin{equation}
\Phi({\mathbf x},z)=\Phi_r({\mathbf x},z)z^{\frac{5}{2}}+\phi_{-1}({\mathbf x})\chi(z)z^{\frac{5}{2}}
+\phi_0({\mathbf x})\chi(z)z^{\frac{5}{2}}\log z+\phi_{1}({\mathbf x})\chi(z)z^{\frac{1}{2}}+\phi_{2}({\mathbf x})z^{-\frac{3}{2}}
 \label{HDE}
\end{equation}
with
\begin{equation}
\left\{
\begin{array}{c}
\phi_{-1}\in L^2(\RR^3_{\mathbf x}),\;\phi_{0}\in H^1(\RR^3_{\mathbf x}),\;\phi_{1},\phi_2\in H^2(\RR^3_{\mathbf x}),\;\;\Phi_r({\mathbf x},z)z^{\frac{5}{2}}\in L^2(\RR^3_{\mathbf x};{\bf h}^4),\\
\nabla_{\mathbf x}\left(\Phi_r({\mathbf x},z)z^{\frac{5}{2}}+\phi_{-1}({\mathbf x})\chi(z)z^{\frac{5}{2}}\right) \in L^2(\RR^3_{\mathbf x};{\bf h}^3),\\
\nabla_{\mathbf x}^2\left(\Phi_r({\mathbf x},z)z^{\frac{5}{2}}+\phi_{-1}({\mathbf x})\chi(z)z^{\frac{5}{2}}+\phi_0({\mathbf x})\chi(z)z^{\frac{5}{2}}\log z\right) \in L^2(\RR^3_{\mathbf x};{\bf h}^2).
\end{array}
\right.
 \label{HEDEDE}
\end{equation}
For convenience and to make more clear the role of the massless graviton, we have omitted the cut-off function $\chi(z)$ in front of $\phi_2({\mathbf x})z^{-\frac{3}{2}}$. It is clear that this minor change does not affect the definition of the spaces since $(1-\chi(z))\phi_2({\mathbf x})z^{-\frac{3}{2}}$ belongs to $H^m(\RR^3_{\mathbf x};{\mathbf h}^4)$ when $\phi_2\in H^m(\RR^3_{\mathbf x})$.

To take account the constraint (\ref{constraint}), we introduce the subspace :
\begin{equation}
{\mathfrak D}_{\alpha}:=\left\{\Phi\in {\mathfrak H}_2;\;\;\phi_{-1}({\mathbf x})
+\alpha_0\phi_0({\mathbf x})+\alpha_1\phi_{1}({\mathbf x})+\alpha_2\phi_{2}({\mathbf x})=0\right\}.
 \label{DALPHA}
\end{equation}

The main result of this paper is the following :
\begin{Theorem}
 \label{theobess}
 Let $\alpha=(\alpha_0,\alpha_1,\alpha_2)$ be in $\RR^3$ satisfying the constraints (\ref{zin}) and (\ref{zinb}). Then for any $\Phi_0\in{\mathfrak H}_1$, 
$\Phi_1\in{\mathfrak H}_0$, the Cauchy problem (\ref{eqad}), (\ref{ciad}) has a unique solution
\begin{equation}
\Phi_{\alpha}\in C^0\left(\RR_t;{\mathfrak H}_1\right)\cap C^1\left(\RR_t;{\mathfrak H}_0\right) \cap C^2\left(\RR_t;{\mathfrak H}_{-1}\right)\cap{\mathcal D}'\left(\RR_t;{\mathfrak D}_{\alpha}\right).
 \label{regadun}
\end{equation}
 Moreover there exists $C,\kappa>0$ independent of $\Phi_j$ such that :
\begin{equation}
\Vert\partial_t\Phi_{\alpha}(t)\Vert_{{\mathfrak H}_0}+\Vert\Phi_{\alpha}(t)\Vert_{{\mathfrak H}_1}\leq C\left(\Vert \Phi_1\Vert_{{\mathfrak H}_0}+\Vert \Phi_0\Vert_{{\mathfrak H}_1}\right)e^{\kappa\mid t\mid},
 \label{estada}
\end{equation}
and for all $\Theta\in C^{\infty}_0(\RR_t)$ we have :
\begin{equation}
\Vert\int\Theta(t)\Phi_{\alpha}(t)dt\Vert_{\mathfrak{H}_2}\leq C \left(\Vert \Phi_1\Vert_{{\mathfrak H}_0}+\Vert \Phi_0\Vert_{{\mathfrak H}_1}\right)\int\ \left(\mid\Theta(t)\mid+\mid\Theta''(t)\mid\right)e^{\kappa\mid t\mid}dt.
 \label{estatdad}
\end{equation}

When $\Phi_0,\Phi_1\in C^{\infty}_0(\RR^3_{\mathbf x}\times ]0,\infty[_z)$, $(\Phi_0,\Phi_1)\neq (0,0)$, then $\phi_2\neq 0$ hence $\Phi_{\alpha}$ is not the Friedrichs solution, moreover $\Phi_{\alpha}\neq\Phi_{\alpha'}$ if $\alpha\neq \alpha'$.
\\

There exists a conserved energy, i.e. a non trivial, continuous quadratic form $\EE_{\alpha}$ defined on ${\mathfrak H}_1\oplus{\mathfrak H}_0$, that satisfies :
\begin{equation}
\forall t\in\RR,\;\;\EE_{\alpha}\left(\Phi_{\alpha}(t),\partial_t\Phi_{\alpha}(t)\right)=\mathcal{E}_{\alpha}(\Phi_0,\Phi_1).
 \label{enrPsi}
\end{equation}
This energy is not positive definite in general but $\EE_{\alpha}$  is equivalent to $\Vert \Phi_0\Vert^2_{\mathfrak H_1}+\Vert \Phi_1\Vert^2_{\mathfrak H_0}$  on $C^{\infty}_0(\RR^3_{\mathbf x}\times]0,\infty[_z)\oplus C^{\infty}_0(\RR^3_{\mathbf x}\times]0,\infty[_z)$ and given for $\Phi_0,\Phi_1\in C^{\infty}_0(\RR^3_{\mathbf x}\times]0,\infty[_z)$ by
\begin{equation}
\begin{split}
\EE_{\alpha}(\Phi_0,\Phi_1)=&\Vert P_1P_2\Phi_0\Vert^2_{L^2}-(\mu_1+\mu_2) \Vert P_2\Phi_0\Vert^2_{L^2}+\mu_1\mu_2 \Vert P_1\Phi_0\Vert^2_{L^2}\\
&+\Vert\nabla_{\mathbf x} P_2\Phi_0\Vert^2_{L^2}-(\mu_1+\mu_2) \Vert\nabla_{\mathbf x}  P_1\Phi_0\Vert^2_{L^2}+\mu_1\mu_2 \Vert \nabla_{\mathbf x} \Phi_0\Vert^2_{L^2}\\
&+\Vert P_2\Phi_1\Vert^2_{L^2}-(\mu_1+\mu_2) \Vert P_1\Phi_1\Vert^2_{L^2}+\mu_1\mu_2 \Vert \Phi_1\Vert^2_{L^2}
\end{split}
 \label{enerexad}
\end{equation}
for some $\mu_1<\mu_2<0$.
When $\alpha$ satisfies (\ref{spev}) or (\ref{spevo}), $\EE_{\alpha}$ is positive  on ${\mathfrak H}_1\oplus{\mathfrak H}_0$.\\

When $\Phi_0\in {\mathfrak D}_{\alpha}$, $\Phi_1\in {\mathfrak H}_1$ then $\Phi_{\alpha}$ satisfies :
\begin{equation}
\Phi_{\alpha}\in C^0\left(\RR_t;{\mathfrak D}_{\alpha}\right)\cap C^1\left(\RR_t;{\mathfrak H}_1\right)\cap C^2\left(\RR_t;{\mathfrak H}_0\right),
 \label{regadeu}
\end{equation}
\begin{equation}
\Vert\partial_t\Phi_{\alpha}(t)\Vert_{{\mathfrak H}_1}+\Vert\Phi_{\alpha}(t)\Vert_{{\mathfrak H}_2}\leq C\left(\Vert \Phi_1\Vert_{{\mathfrak H}_1}+\Vert \Phi_0\Vert_{{\mathfrak H}_2}\right)e^{\kappa\mid t\mid}.\\
 \label{estadade}
\end{equation}

There exists $M>0$ such that if $\hat{\Phi}_j(\pmb\xi,z)=0$ for all $\mid\pmb\xi\mid\leq M$, then we can take $\kappa=0$ in the estimates (\ref{estada}), (\ref{estatdad}) and (\ref{estadade}) and $\EE_{\alpha}(\Phi_0,\Phi_1)>0$.\\

When the equation
\begin{equation}
\log x+2\alpha_0+\frac{8\alpha_1}{x}-\frac{32\alpha_2}{x^2}=0,
 \label{funclamads}
\end{equation}
has a solution $x=m^2$, $m>0$, then $\phi_{[m]}(t,{\mathbf x})z^{\frac{1}{2}}K_2(mz)$ where $\phi_{[m]}\in C^0(\RR_t;H^2(\RR^3_{\mathbf x}))\cap C^1(\RR_t;H^1(\RR^3_{\mathbf x}))$ is a solution of $\partial_t^2\phi_{[m]}-\Delta_{\mathbf x}\phi_{[m]}-m^2\phi_{[m]}=0$, is  a solution that  satisfies (\ref{regadeu}).\\

When $\alpha$ satisfies (\ref{spevo}), the massless graviton $\Phi_G(t,{\mathbf x},z):=\phi_{[0]}(t,{\mathbf x})z^{-\frac{3}{2}}$ where $\phi_{[0]}\in C^0(\RR_t;H^2(\RR^3_{\mathbf x}))$ is solution of $\partial^2_t\phi_{[0]}-\Delta_{\mathbf x}\phi_{[0]}=0$, is a solution of (\ref{eqad}) that satisfies (\ref{regadeu}), and its energy is given by
\begin{equation}
\EE_{\alpha}(\Phi_G,\partial_t\Phi_G)=\Vert z^{-\frac{3}{2}}\Vert_0^2\int_{\RR^3_{\mathbf x}}\vert\nabla_{t,\mathbf x}\phi_{[0]}(t,\mathbf x)\vert^2d\mathbf{x}.
 \label{enegrav}
\end{equation}
\end{Theorem}

{\it Proof of Theorem \ref{theobess}.} We shall use the partial Fourier transform with respect to ${\mathbf x}$ that is denoted ${\mathcal F}_{\mathbf x}$. Let $\Phi_{\alpha}$ be a solution of (\ref{eqad}), (\ref{ciad}), (\ref{regadeu}). Given $T>0$, $\Phi_{\alpha}\in H^1(]-T,T[;{\mathfrak H}_1)\subset L^2(\RR^3_{\mathbf x}; H^1(]-T,T[;{\mathfrak h}_1))$. Then ${\mathcal F}_{\mathbf x}\Phi_{\alpha}\in  L^2(\RR^3_{\pmb \xi}; H^1(]-T,T[;{\mathfrak h}_1))\subset  L^2(\RR^3_{\pmb \xi}; C^0([-T,T];{\mathfrak h}_1))$. We have also $\partial_t\Phi_{\alpha}\in H^1(]-T,T[;{\mathfrak H}_0)\subset L^2(\RR^3_{\mathbf x}; H^1(]-T,T[;{\mathfrak h}_0))$. Then $\partial_t{\mathcal F}_{\mathbf x}\Phi_{\alpha}\in  L^2(\RR^3_{\pmb \xi}; H^1(]-T,T[;{\mathfrak h}_0))\subset  L^2(\RR^3_{\pmb \xi}; C^0([-T,T];{\mathfrak h}_0))$. Moreover $\Phi_{\alpha}\in L^2(]-T,T[;{\mathfrak D}_{\alpha})\subset L^2(\RR^3_{\mathbf x}; L^2(]-T,T[;{\mathfrak d}_{\alpha}))$. Then ${\mathcal F}_{\mathbf x}\Phi_{\alpha}\in  L^2(\RR^3_{\pmb \xi}; L^2(]-T,T[;{\mathfrak d}_{\alpha}))$.

We deduce that for almost all $\pmb \xi\in\RR^3$,  ${\mathcal F}_{\mathbf x}\Phi_{\alpha}(t,{\pmb\xi}, z)$ is the unique solution $\psi_{\pmb\xi}$, satisfying (\ref{reglagla}), of (\ref{zepbun}) with 
\begin{equation}
m=\mid\pmb\xi\mid,\;\;f(z)= {\mathcal F}_{\mathbf x}\Phi_{0}({\pmb\xi}, z),\;\;g(z)={\mathcal F}_{\mathbf x}\Phi_{1}({\pmb\xi}, z).
 \label{cliffi}
\end{equation}
Hence we conclude that
\begin{equation}
\Phi_{\alpha}(t,{\mathbf x}, z)={\mathcal F}_{\pmb \xi}^{-1}\left(\psi_{\pmb\xi}(t,z)\right)({\mathbf x}),
 \label{fanala}
\end{equation}
and we get the uniqueness of the solution.

More generally, when $\Phi_{\alpha}$ is a solution of (\ref{eqad}), (\ref{ciad}), (\ref{regadun}), we take $\theta\in C^{\infty}_0(\RR)$ such that $0\leq\theta$, $\int\theta(t)dt=1$, and we consider $\Phi_{\alpha, n}(t,{\mathbf x},z)=n\int\theta(nt-ns) \Phi_{\alpha}(s,{\mathbf x},z)ds$. We can easily prove that $\Phi_{\alpha, n}$ tends to $\Phi_{\alpha}$ in $C^0\left(\RR_t;{\mathfrak H}_1\right)\cap C^1\left(\RR_t;{\mathfrak H}_0\right) \cap C^2\left(\RR_t;{\mathfrak H}_{-1}\right)\cap{\mathcal D}'\left(\RR_t;{\mathfrak D}_{\alpha}\right)$ as $n\rightarrow\infty$, and $\Phi_{\alpha, n}$ is a solution of (\ref{eqad}), (\ref{regadeu}). The previous result shows that
$$
\Phi_{\alpha,n}(t,{\mathbf x}, z)={\mathcal F}_{\pmb \xi}^{-1}\left(\psi_{\pmb\xi, n}(t,z)\right)({\mathbf x}),
$$
where $\psi_{\pmb\xi, n}$ is solution of (\ref{zepbun}) with $m=\mid\pmb\xi\mid$, $f(z)= {\mathcal F}_{\mathbf x}\Phi_{\alpha,n}(0,{\pmb\xi}, z)$, $g(z)={\mathcal F}_{\mathbf x}\partial_t\Phi_{\alpha,n}(0,{\pmb\xi}, z)$ satisfying (\ref{reglagla}). Since $\Phi_{\alpha,n}(0,{\mathbf x}, z)$ and  $\partial_t\Phi_{\alpha,n}(0,{\mathbf x}, z)$ tend respectively to $\Phi_0$ and $\Phi_1$ in ${\mathfrak H}_1$ and ${\mathfrak H}_0$, then ${\mathcal F}_{\mathbf x}\Phi_{\alpha,n}(0,{\pmb\xi}, z)$ and ${\mathcal F}_{\mathbf x}\partial_t\Phi_{\alpha,n}(0,{\pmb\xi}, z)$ tend respectively to ${\mathcal F}_{\mathbf x}\Phi_{0}({\pmb\xi}, z)$ and ${\mathcal F}_{\mathbf x}\Phi_{1}({\pmb\xi}, z)$ in $L^2(\RR^3_{\pmb\xi};{\mathfrak h}_1)$ and  $L^2(\RR^3_{\pmb\xi};{\mathfrak h}_0)$. We deduce by (\ref{contipsi}) that $\psi_{\pmb\xi, n}$ tends in $L^2(\RR^3_{\pmb\xi};L^2([-T,T];{\mathfrak h}_1)$ to the solution $\psi_{\pmb\xi}$ of (\ref{zepbun}),  (\ref{contipsi})  with the data (\ref{contipsi}). We conclude that (\ref{fanala}) is true again and the proof of the uniqueness is complete.

To establish the existence of the solution, it is sufficient to solve the Cauchy problem and to get estimates (\ref{estada}), (\ref{estada}), (\ref{estadade}) for a dense subspace of initial data. Hence we consider the case where there exists $R>0$ such that ${\mathcal F}_{\mathbf x}\Phi_{j}({\pmb\xi}, z)=0$ for any $\mid\pmb\xi\mid>R$. Then we get by the Lebesgue theorems, the Parseval equality and Theorem \ref{theobess}, that
$$
\Phi_{\alpha}(t,{\mathrm x},z):=\frac{1}{(2\pi)^{\frac{3}{2}}}\int_{\mid\pmb\xi\mid\leq R}e^{i{\mathbf x}.{\pmb\xi}}\psi_{\pmb\xi}(t,z)d{\pmb\xi}
$$
is the wanted solution, moreover estimates (\ref{estada}), (\ref{estatdad}), (\ref{estadade}) directly follow from the integration of (\ref{contipsi}),  (\ref{contipsid}), (\ref{contipside}) with respect to $\pmb\xi$, and we can take $\kappa=0$ when $\hat{\Phi}_j({\pmb\xi},z)=0$ for all $\mid\pmb\xi\mid\leq M$  where $M=\sqrt{K}$.

If $\phi_2=0$ for $\Phi_0,\Phi_1\in C^{\infty}_0(\RR^3_{\mathbf x}\times ]0,\infty[_z)$,  then we have ${\mathcal F}_{\mathbf x}\phi_2$ and $\psi_{\pmb\xi}(t,z)$ is a Friedrichs solution. The Theorem  \ref{theobess} implies that $\Phi_0=\Phi_1=0$. Now if $\Phi_{\alpha}=\Phi_{\alpha'}$, then   ${\mathcal F}_{\mathbf x}\Phi_{\alpha}={\mathcal F}_{\mathbf x}\Phi_{\alpha'}$ and this theorem assures that $\alpha = \alpha'$.

The properties of the energy are obtained by the same way from (\ref{enrpsi}) and  (\ref{enerex}) with the Parseval equality and the formula 
$$
\EE_{\alpha}\left(\Phi_{\alpha}(t),\partial_t\Phi_{\alpha}(t)\right)=\int \mathcal{E}_{\alpha}\left(\psi_{\pmb\xi}(t),\partial_t\psi_{\pmb\xi}(t)\right) d{\pmb\xi}.
$$
We also have with (\ref{pschit}) :
$$
\EE_{\alpha}\left(\Phi_0,\Phi_1\right)=\int\left<{\mathbf A}_{\alpha}{\mathcal F}_{\mathbf x}\Phi_{0}({\pmb\xi}, .); {\mathcal F}_{\mathbf x}\Phi_{0}({\pmb\xi}, .)\right>_0d{\pmb\xi}+
\Vert\nabla_{\mathbf x}\Phi_0\Vert^2_{\mathfrak{H}_0}+\Vert\Phi_1\Vert^2_{\mathfrak{H}_0},
$$
that proves (\ref{enegrav}).
Finally, since the proposition \ref{propbessel} assures that for $m>0$, $\psi_m(z):=z^{\frac{1}{2}}K_2(mz)$, and for $m=0$ $\psi_0(z):=z^{-\frac{3}{2}}$, satisfy $ (P_2+m^2)\psi_m=0$, and belong to ${\mathfrak{d}}_{\alpha}$ when $x=m^2>0$ is solution of (\ref{funclamads}), or $\alpha$ satisfies (\ref{spevo}) for $m=0$. We conclude that $\Phi_{\alpha}(t,{\mathbf x},z)=\phi_{[m]}(t,{\mathbf x})\psi_m(z)$ are solutions of (\ref{eqad}) satisfying (\ref{regadeu}).

\fin

We achieve this paper by some comments. If we expand the strong solution as
$$
\Phi_{\alpha}(t,{\mathbf x} ,z)=\phi_r (t,{\mathbf x} ,z)+\phi_0(t,{\mathbf x})\chi(z)z^{\frac{5}{2}}\log z+\phi_1 (t,{\mathbf x})\chi(z)z^{\frac{1}{2}}+\phi_2 (t,{\mathbf x})z^{-\frac{3}{2}}
$$
then we can see with (\ref{HDE}) and (\ref{HEDEDE}) that the equation (\ref{eqad}) is equivalent to a system of coupled PDEs (we denote $\Box :=\partial_t^2-\Delta_{\mathbf x}$) :
$$
\Box\phi_2+4\phi_1=0,
$$
$$
\Box\phi_1-4\phi_0=0,
$$
\begin{equation*}
\begin{split}
\left[\Box -\partial_z^2+\frac{15}{4z^2}\right]&\left(\phi_r+\chi(z)z^{\frac{5}{2}}\log(z)\phi_0\right)\\
&=-4\chi(z) z^{\frac{1}{2}}\phi_0+\left(\chi''(z)z^{\frac{1}{2}}+\chi'(z)z^{-\frac{1}{2}}+4(1-\chi(z))z^{-\frac{3}{2}}\right)\phi_1,
\end{split}
\end{equation*}
supplemented by the boundary contraint at the time-like horizon :
$$
\lim_{z\rightarrow 0}z^{-\frac{5}{2}}\phi_r(t,\mathbf{x},z)+\alpha_0\phi_0(t,\mathbf{x})+\alpha_1\phi_1(t,\mathbf{x}) +\alpha_2\phi_2(t,\mathbf{x})=0.
$$
We note that $\phi_2$ is not a free wave in the Minkowski space-time (see below for a link with the massless graviton), and $\Phi_F:=\phi_r+\chi(z)z^{\frac{5}{2}}\log(z)\phi_0$ is a Friedrichs solution of the inhomogeneous wave equation of the gravitational fluctuations, {\it i.e.} $\Phi_F$ satisfies (\ref{enerdir}).

A particularly significant family of constraints on the boundary of the Anti-de Sitter universe is given by the condition (\ref{spevo}) that corresponds to 
$$
\phi_{-1}(t,{\mathbf x})
+\alpha_0\phi_0(t,{\mathbf x})+\alpha_1\phi_{1}(t,{\mathbf x})=0
$$
with
$$
0<\alpha_1,\;\;-\frac{1}{2}-\frac{3}{2}\log2<\alpha_0+\frac{1}{2}\log\alpha_1<\frac{1}{4}-\frac{1}{2}\log2-\gamma.
$$
In this case the energy is positive and $\sqrt{\EE_{\alpha}(\Phi_0,\Phi_1)}$ is a norm on  $\mathfrak{H}_1\times\mathfrak{H}_0$. Hence we can consider the Hilbert space $\mathfrak{K}_1\times\mathfrak{H}_0$ defined as  the completion of  this space for this norm. We remark that   $\mathfrak{H}_1\neq  \mathfrak{K}_1$ since
$$
\Vert\phi(\mathbf{x})z^{-\frac{3}{2}}\Vert_{\mathfrak{K}_1}^2=\Vert z^{-\frac{3}{2}}\Vert_0^2\int_{\RR^3_{\mathbf x}}\vert\nabla_{\mathbf x}\phi(\mathbf x)\vert^2d\mathbf{x},\;\;\Vert\phi(\mathbf{x})z^{-\frac{3}{2}}\Vert_{\mathfrak{H}_1}^2=\Vert\phi(\mathbf{x})z^{-\frac{3}{2}}\Vert_{\mathfrak{K}_1}^2+\Vert z^{-\frac{3}{2}}\Vert_{\mathfrak{h}_1}^2\int_{\RR^3_{\mathbf x}}\vert\phi(\mathbf x)\vert^2d\mathbf{x}.
$$
Then the Cauchy problem is well posed in $\mathfrak{K}_1\times\mathfrak{H}_0$ and the solution is given by a unitary group. Finally (\ref{gravexpand}), (\ref{graviton}) and (\ref{fanala}) allow to split the solution $\Phi_{\alpha}$  into a massless graviton $\Phi_G$ and an orthogonal part $\Phi^{\perp}$, solutions of (\ref{eqad}) satisfying :
$$
\Phi_{\alpha}=\Phi_G+\Phi^{\perp},\;\;\Phi_G(t,\mathbf{x},z)=\phi_{[0]}(t,\mathbf{x})z^{-\frac{3}{2}},
$$
where
$$
\partial_t^2 \phi_{[0]}-\Delta_{\mathbf{x}}\phi_{[0]}=0,\;\;\phi_{[0]}(0,\mathbf{x})=\Vert z^{-\frac{3}{2}}\Vert_0^{-2}\left<\Phi_0(\mathbf{x},.);z^{-\frac{3}{2}}\right>_0,\;\;
\partial_t\phi_{[0]}(0,\mathbf{x})=\Vert z^{-\frac{3}{2}}\Vert_0^{-2}\left<\Phi_1(\mathbf{x},.);z^{-\frac{3}{2}}\right>_0,
$$
and
for all $t\in\RR$ and almost $\mathbf{x}\in\RR^3$,
$$
\left<\Phi^{\perp}(t,\mathbf{x},.);z^{-\frac{3}{2}}\right>_0=0.
$$
In the spirit of (\ref{scatgr}), we conjecture that
$$
\lim_{\mid t\mid\rightarrow\infty}\Vert \nabla_{t,\mathbf{x}}\phi_{[0]}(t,.)-\nabla_{t,\mathbf{x}}\phi_2(t,.)\Vert_{L^2(\RR^3_{\mathbf x})}=0,
$$
that is to say, the more singular part of the gravitational wave is asymptotically given by the massless graviton. Last but not least, we let open the deep question on the ``true'' constraint on the boundary on the Anti-de Sitter universe, among the large family of the boundary conditions that we have introduced in this work.

\end{document}